\documentclass[aps,prd,twocolumn,superscriptaddress,showkeys]{revtex4}
\usepackage{amssymb,amsmath,graphicx,enumerate}
\usepackage{subfigure}

\usepackage{dcolumn,booktabs}
\usepackage{longtable}
\usepackage[dvipdfm,colorlinks=true,citecolor=blue,pdfstartview=FitH]{hyperref}

\def\bea{\begin{equation}}
\def\eea{\end{equation}}
\newcommand{\rt}{Regge trajectory}
\newcommand{\rts}{Regge trajectories}
\newcommand{\bfr}{{\bf r}}

\newcommand{\bfpa}{{|\bf p|}}

\newcommand{\alp}{\ensuremath{\alpha^\prime}}
\newcommand{\sse}{spinless Salpeter equation}
\newcommand{\nr}{nonrelativistic}
\newcommand{\ur}{ultrarelativistic}

\begin{document}
\title{Revisiting the pion Regge trajectories}
\author{Jiao-Kai Chen}
\email{chenjk@sxnu.edu.cn, chenjkphy@outlook.com}
\affiliation{School of Physics and Information Engineering, Shanxi Normal University, Taiyuan 030031, China}
\date{\today}

\begin{abstract}
We propose a model-independent ansatz $M={\beta_x}\left(x+c_0\right)^{\nu}+c_1$ ($x=l,\,n_r$) and then use it to fit the orbital and radial pion Regge trajectories without the preset values. It is shown that nonzero $c_1$ is reasonable and acceptable. Nonzero $c_1$ gives an explanation for the nonlinearity of the pion Regge trajectories in the usually employed $(M^2,\,x)$ plane. As $m_R$ or $c_1$ is chosen appropriately, both the orbital and radial pion Regge trajectories are linear in the $((M-m_R)^2,\,x)$ plane whether the $\pi^0$ is included or not on the Regge trajectories. The fitted pion Regge trajectories suggest $0.45\le\nu\le0.5$, which indicates the confining potential $r^a$ with $9/11{\le}a\le1$. Moreover, it is illustrated in the appendix B that $m_R$ can be nonzero for the light nonstrange mesons. We present discussions in the appendix A on the structure of the Regge trajectories plotted in the $(M,\,x)$ plane and on the structure of the Regge trajectories in the $((M-m_R)^2,\,x)$ plane based on the potential models and the string models.
\end{abstract}


\maketitle


\section{Introduction}

\begin{figure}[!phtb]
\centering
\subfigure[]{\label{subfigure:fiterr}\includegraphics[scale=0.85]{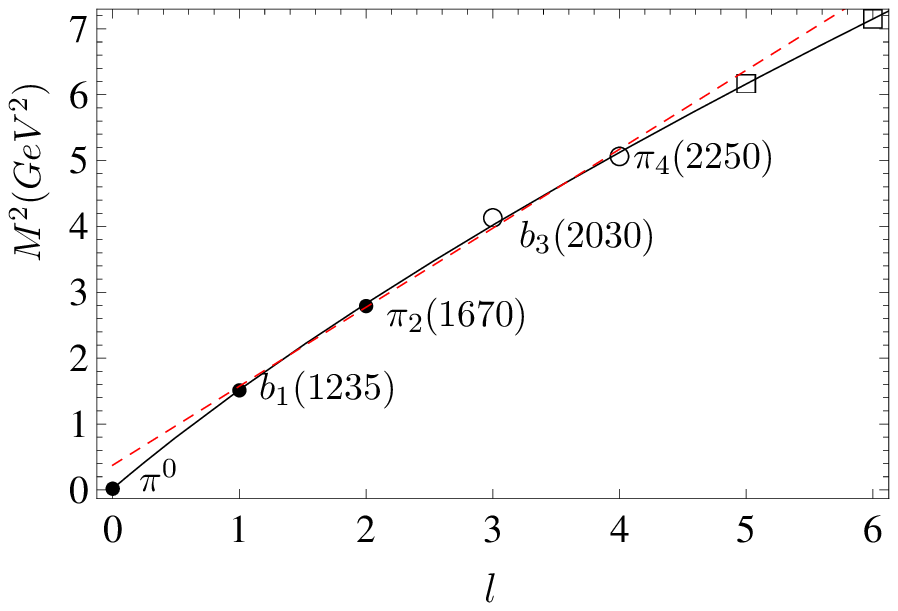}}
\subfigure[]{\label{subfigure:fitbeta}\includegraphics[scale=0.85]{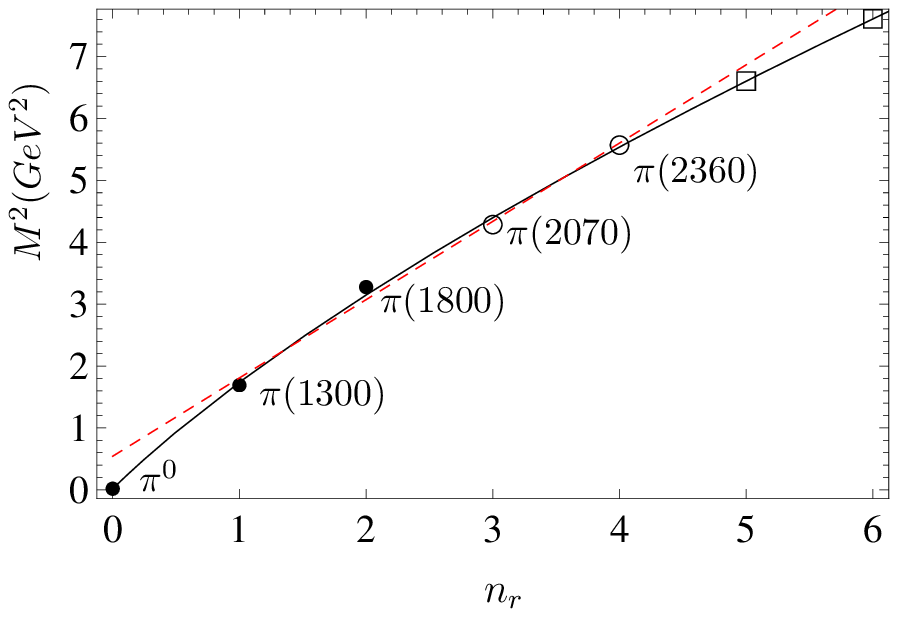}}
\caption{The fitted orbital and radial {\rts} for pion, which are plotted in the $(M^2,\,l)$ plane and in the $(M^2,\,n_r)$ plane, respectively. These figures are from Ref. \cite{Chen:2021kfw}. The well-established states are given by solid dots and the unwell-established states are given by circles. Open squares are predicted masses.}\label{fig:regm2old}
\end{figure}

Understanding the spectrum of hadrons reveals information on the non-perturbative aspects of QCD \cite{Godfrey:1998pd} and on the inner structure of hadrons.
The {\rt} is one of the effective approaches for studying hadron spectra \cite{Regge:1959mz,Collins:1971ff,Chew:1961ev,Hey:1982aj,Inopin:2001ub,Chen:2016spr,
Chen:2016qju,Inopin:1999nf,Chew:1962eu,Brodsky:2017qno}.
The orbital and radial {\rts} \cite{note} for pion are often taken as being approximately linear in the $(M^2,\,l)$ plane and in the $(M^2,\,n_r)$ plane, respectively, $M^2={\alpha_1}l+{\alpha_2}n_r+c$ \cite{Anisovich:2000kxa,Afonin:2006vi,Shifman:2007xn,Ebert:2009ub,Masjuan:2012gc,
Kobzarev:1992wt,Lucha:1991ce,Afonin:2021cwo}, where $M$ is the mass, $\alpha_1$ and $\alpha_2$ are the Regge slopes, $l$ is the orbital angular momentum, $n_r$ is the radial quantum number, and $c$ is a constant.
The pion {\rts} are in fact nonlinear when they are examined more precisely, see Fig. \ref{fig:regm2old}.
Many authors have discussed the nonlinearity of the pion {\rts}. In Ref. \cite{Tang:2000tb}, the authors note that the orbital pion {\rt} is nonlinear by the "zone test", $M^2=-0.1077J^2+1.6003J+0.019$, where $J$ is the total angular momentum. In Ref. \cite{Brisudova:1999ut}, the pion {\rt} is discussed by using the square-root trajectory. In Ref. \cite{Sharov:2013tga}, the author gives the nonlinear pion orbital {\rt} with corrections based on the string model. In Ref. \cite{Sonnenschein:2014jwa}, the pion {\rt} is nonlinear as the masses of quarks are considered. In Refs. \cite{Chen:2021kfw,Chen:2018hnx,Chen:2018bbr}, the pion {\rts} are fitted by a nonlinear formula, $M^2=2.78(0.8+n_r)^{2/3}-2.38$.
It is known that the significant nonlinearity of the {\rts} for the heavy mesons arises from the nonrelativisity of heavy mesons \cite{Sergeenko:1994ck,Inopin:1999nf,Sonnenschein:2018fph,Sonnenschein:2014jwa,
Cotugno:2009ys,Burns:2010qq,Kruczenski:2004me,MartinContreras:2020cyg,Chen:2021kfw,Chen:2018hnx,
Chen:2018nnr,Chen:2018bbr} due to the heavy masses of quarks. In this work, we revisit the pion {\rts} and present discussions on the nonlinearity of them.

This paper is organized as follows: In Sec. \ref{sec:fit}, we fit the orbital and radial pion {\rts} by using four points and five points on the {\rts}, respectively. The discussions are presented in Sec. \ref{sec:dis} and the conclusions are in Sec. \ref{sec:conclusions}.

\section{Fit of the pion {\rts}}\label{sec:fit}
In this section, we fit the orbital and radial {\rts} for the pion by employing a newly proposed ansatz. The ansatz is model-independent and therefore the fit is model-independent. Then we obtain the fitted parameters without the preset values.

\subsection{Preliminaries}

The {\rts} can be written in different forms, such as $l=l(M)$,  $M^2=f(l,\,n_r)$, $M=M(l,\,n_r)$ \cite{Burns:2010qq}, $(M-m)^2=g(l,\,n_r)$ \cite{Afonin:2020bqc,Chen:2017fcs,Jia:2019bkr} and so on.
We use the following ansatz inspired by Refs. \cite{Chen:2021kfw,MartinContreras:2020cyg,Brau:2000st}
\bea\label{grform}
M={\beta_x}\left(x+c_0\right)^{\nu}+c_1\;\;(x=l,\,n_r)
\eea
to fit the orbital and radial {\rts} \cite{note} for the pion in the $(M,\,x)$ plane. $\beta_x$ is the slope. The constants $c_0$ and $c_1$ vary with different {\rts}. The exponent $\nu$ which relates to the dynamics of mesons is regarded as a free parameter in fit.
As $0<\nu<1$, $\nu=1$ or $\nu>1$, Eq. (\ref{grform}) becomes concave downwards, linear or convex upwards, respectively. $\nu\in(0,1)$ is used to find the appropriate value because the pion orbital and radial {\rts} are obviously concave in the $(M,\,n_r)$ and $(M,\,l)$ planes, see Fig. \ref{fig:regm}.

The used data are listed in the 3rd column in Table \ref{tab:fitmass}.
The quality of a fit is measured by the quantity $\chi^2$ defined by \cite{Sonnenschein:2014jwa}
\bea\label{chi2}
\chi^2=\frac{1}{N-1}\sum_{i=1}^N\left(\frac{M_{fi}-M_{ei}}{M_{ei}}\right)^2,
\eea
where $N$ is the number of points on the trajectory, $M_{fi}$ is the fitted value and $M_{ei}$ is the experimental value of the $i$-th particle.

The fit is calculated by using MATHEMATICA program. $c_0$, $c_1$ and $\nu$ are free parameters. $\beta_x$ is calculated by using the FindFit function and Eq. (\ref{grform}). The quantity $\chi^2$ is calculated by using Eq. (\ref{chi2}). By minimizing the $\chi^2$ quantity for a given $\nu$, the parameters $c_0$, $c_1$ and $\beta_x$ are determined for each value of $\nu$.

\begin{table*}[!phtb]
\caption{Mesons' masses (in units of ${\rm MeV}$). The experimental masses are from PDG \cite{ParticleDataGroup:2020ssz}. The fitted results by using the {\rts} [Eq. (\ref{grform})] are shown in comparison with the theoretical values in the EFG model \cite{Ebert:2009ub} and in the GI model \cite{Godfrey:1985xj}. Fit5 are obtained by using five points on the {\rts} while Fit4 are fitted by excluding $\pi^0$. A question mark ($?$) indicates an unwell-established state. The fitted {\rts} are listed in Table \ref{tab:fitreg}.}
\label{tab:fitmass}
\centering
\begin{tabular*}{\textwidth}{@{\extracolsep{\fill}}ccllllllrll@{}}
\hline\hline
State      & Meson           & PDG \cite{ParticleDataGroup:2020ssz}     & EFG \cite{Ebert:2009ub}   & GI \cite{Godfrey:1985xj}   & \multicolumn{3}{c}{Fit5}  & \multicolumn{3}{c}{Fit4}\\
           &                 &                                   &       &                    & $\nu=0.4$   & $\nu=0.45$  & $\nu=0.5$
& $\nu=0.4$   & $\nu=0.45$  & $\nu=0.5$\\
\hline
$1^1S_0$   & $\pi^0$             & $134.9768\pm0.0005$          &154 &150
  & 135  & 136  & 135     & $-$115  & 90 & 255 \\

$2^1S_0$   & $\pi(1300)$         & $1300\pm100$                 &1292 & 1300
  & 1307  & 1302  & 1269     & 1310    & 1312 & 1315 \\

$3^1S_0$   & $\pi(1800)$         & $1810^{+9}_{-11}$             &1788 &1880
  & 1764  & 1756  & 1739     & 1765    & 1759  & 1754 \\

$4^1S_0$   & $\pi(2070)$?        & $2070\pm35$                    &2073 &
  & 2096  & 2095  & 2099     & 2096    & 2093  & 2091 \\

$5^1S_0$   & $\pi(2360)$?         & $2360\pm25$                   &2385 &
  & 2368  & 2376  & 2403     & 2366    & 2370  & 2375 \\

$6^1S_0$   &                      &                               &                &
  & 2601  & 2621  & 2671     & 2598    & 2611  & 2625  \\

$7^1S_0$   &                      &                               &                &
  & 2808  & 2840  & 2913     & 2803    & 2827  & 2851  \\

$8^1S_0$   &                      &                               &                &
  & 2995  & 3039  & 3135     & 2989    & 3023  & 3060 \\
\hline

$1^1S_0$   & $\pi^0$              & $134.9768\pm0.0005$          &154 &150
   & 135   & 135  & 135   & $-$175  & 25   & 185 \\

$1^1P_1$   & $b_1(1235)$           & $1229.5\pm3.2$              &1258 &1220
   & 1226   & 1235  & 1213   & 1229    & 1231  & 1232 \\

$1^1D_2$   & $\pi_2(1670)$        & $1670.6^{+2.9}_{-1.2}$        &1643 &1680
   & 1675   & 1673  & 1660   & 1678    & 1672  & 1666  \\

$1^1F_3$   & $b_3(2030)$?         & $2032\pm12$                   &1884 &2030
   & 2004   & 2000  & 2002   & 2004    & 2002  & 1998  \\

$1^1G_4$   & $\pi_4(2250)$?       & $2250\pm15$                   & 2092 &2330
   & 2272   & 2272  & 2291   & 2270    & 2275  & 2279 \\

$1^1H_5$   &                      &                                &                &
   & 2502   & 2508  & 2545   & 2498    & 2513  & 2526  \\

$1^1I_6$   &                      &                                &                &
   & 2707   & 2719  & 2776   & 2700    & 2726  & 2750 \\
\hline\hline
\end{tabular*}
\end{table*}

\begin{table*}[!phtb]
\caption{The fitted pion {\rts} (in units of ${\rm GeV}$) by using four points (Fit4) and by using five points (Fit5). }
\centering
\begin{tabular*}{\textwidth}{@{\extracolsep{\fill}}cllclc@{}}
\hline\hline
               &                & \multicolumn{2}{c}{Fit4}                 & \multicolumn{2}{c}{Fit5}\\
               &                & Traj.                     & $\chi^2$                   &   Traj.         & $\chi^2$  \\
\hline
               & $\nu=0.4$        &$M=1.404\,l^{0.4}-0.175$   & $9.5\times10^{-5}$                        &$M=1.424(l+0.03)^{0.4}-0.215$     & $7.7\times10^{-5}$\\
Orbital Traj.  & $\nu=0.45$      &$M=1.206\,l^{0.45}+0.025$     & $1.15\times10^{-4}$                         & $M=1.198(l+0.004)^{0.45}+0.035$     & $9.0\times10^{-5}$\\
               & $\nu=0.5$        & $M=1.047\,l^{0.5}+0.185$     & $1.5\times10^{-4}$                         & $M=1.078\,l^{0.5}+0.135$      & $1.9\times10^{-4}$ \\
\hline
               & $\nu=0.4$        &$M=1.425\,n_r^{0.4}-0.115$     & $2.8\times10^{-4}$                        & $M=1.438(n_r+0.016)^{0.4}-0.14$      & $2.2\times10^{-4}$ \\
Radial Traj.   & $\nu=0.45$      & $M=1.222\,n_r^{0.45}+0.09$     & $3.4\times10^{-4}$
                 & $M=1.241(n_r+0.002)^{0.45}+0.06$      & $2.8\times10^{-4}$ \\
               & $\nu=0.5$        & $M=1.060\,n_r^{0.5}+0.255$    & $4.1\times10^{-4}$                         & $M=1.134\,n_r^{0.5}+0.135$         & $6.7\times10^{-4}$ \\
\hline\hline
\end{tabular*}\label{tab:fitreg}
\end{table*}

\begin{figure}[!phtb]
\centering
\subfigure[]{\label{subfig:fiterr}\includegraphics[scale=0.74]{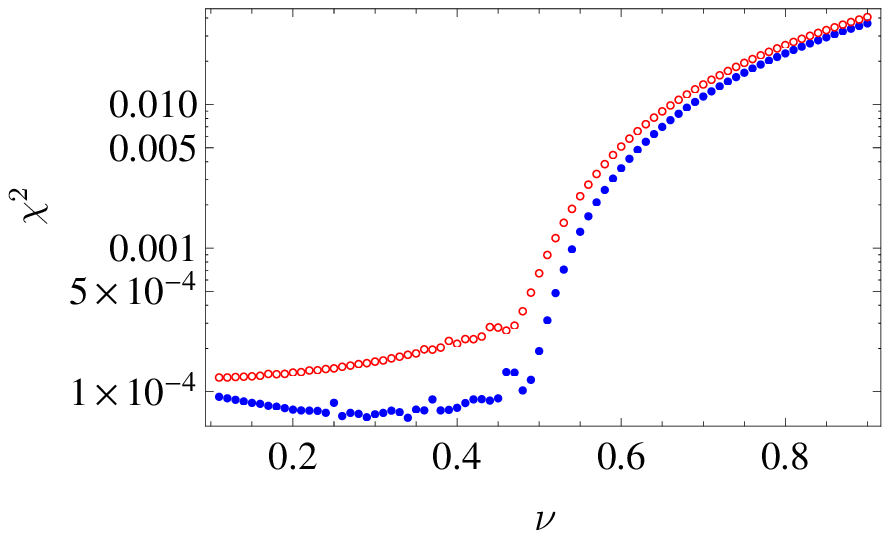}}
\subfigure[]{\label{subfig:fitc0}\includegraphics[scale=0.7]{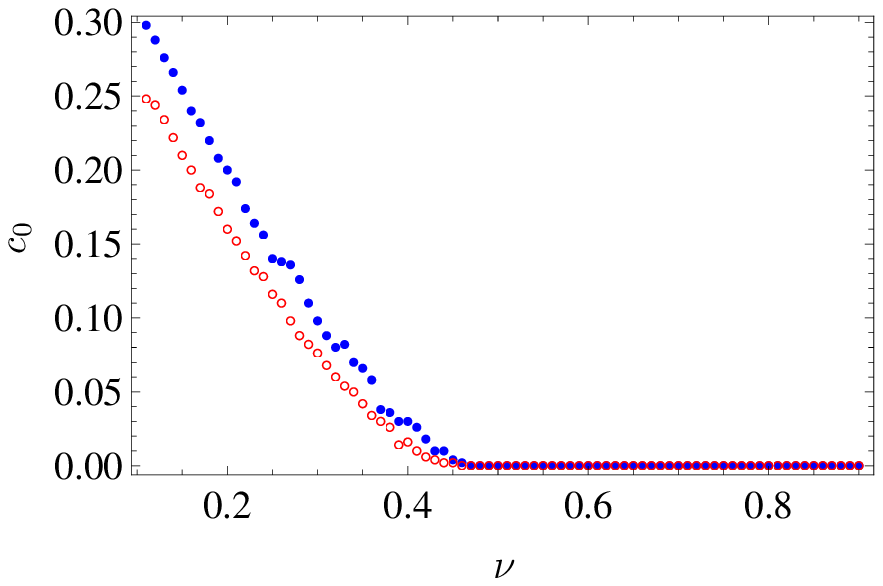}}
\subfigure[]{\label{subfig:fitc1}\includegraphics[scale=0.7]{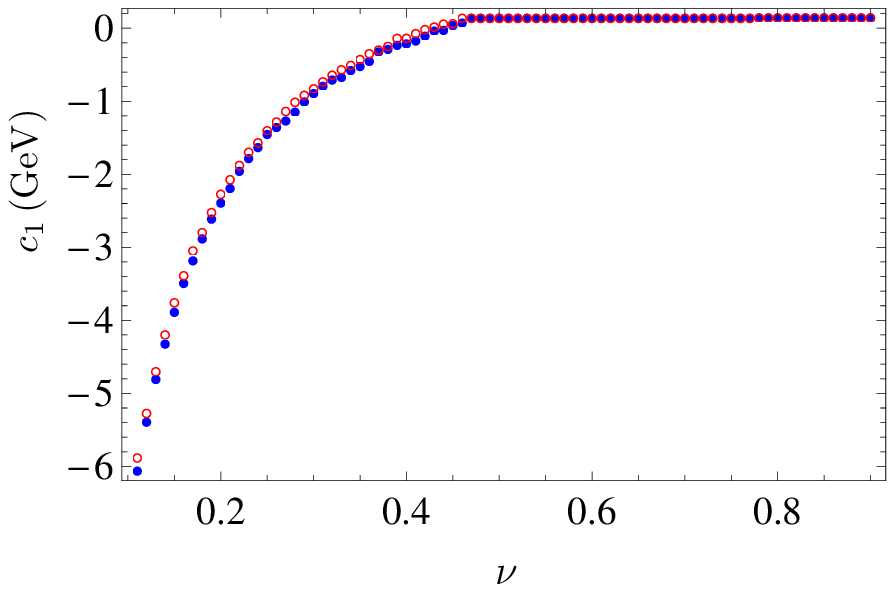}}
\subfigure[]{\label{subfig:fitbeta}\includegraphics[scale=0.7]{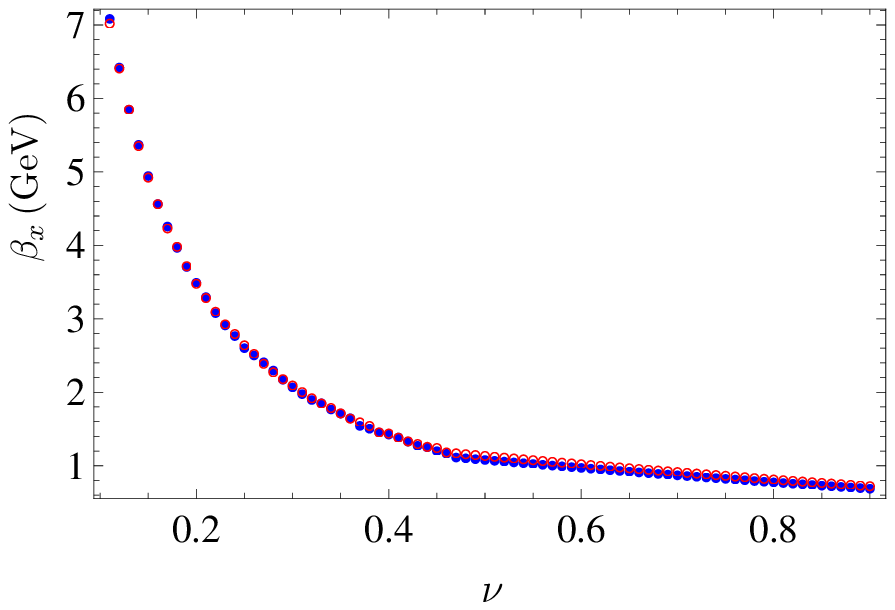}}
\caption{The quantities $\chi^2$, $c_1$, $c_0$ and $\beta_x$ ($x=l,\,n_r$) vary with different exponent $\nu$. The orbital {\rt} (the blue dots) and the radial {\rt} (the red circles) are fitted by using five points on the {\rts}, that is, $\pi^0$ is included in fit. The used formulas are Eqs. (\ref{grform}) and (\ref{chi2}).}\label{fig:fit5}
\end{figure}

\begin{figure}[!phtb]
\centering
\subfigure[]{\label{subfigure:fit4err}\includegraphics[scale=0.74]{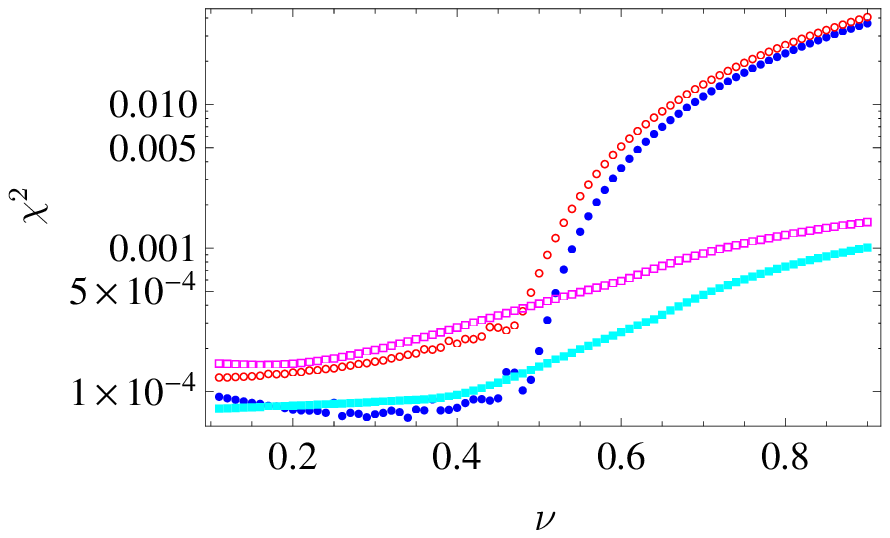}}
\subfigure[]{\label{subfigure:fit4c0}\includegraphics[scale=0.7]{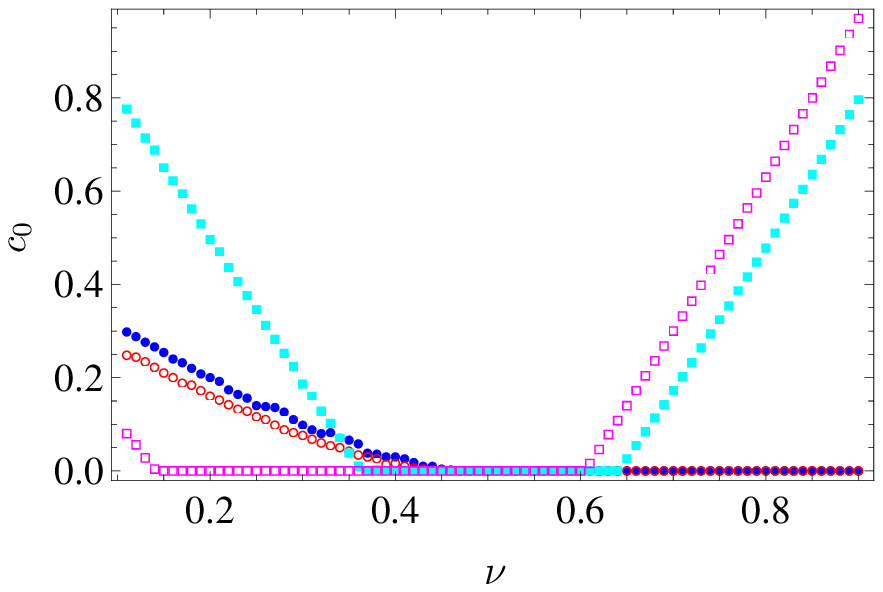}}
\subfigure[]{\label{subfigure:fit4c1}\includegraphics[scale=0.7]{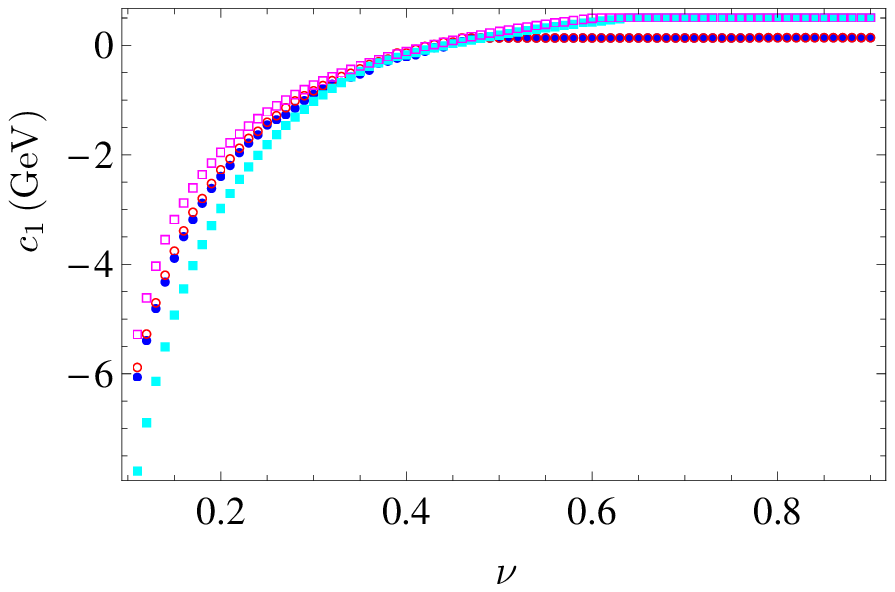}}
\subfigure[]{\label{subfigure:fit4beta}\includegraphics[scale=0.7]{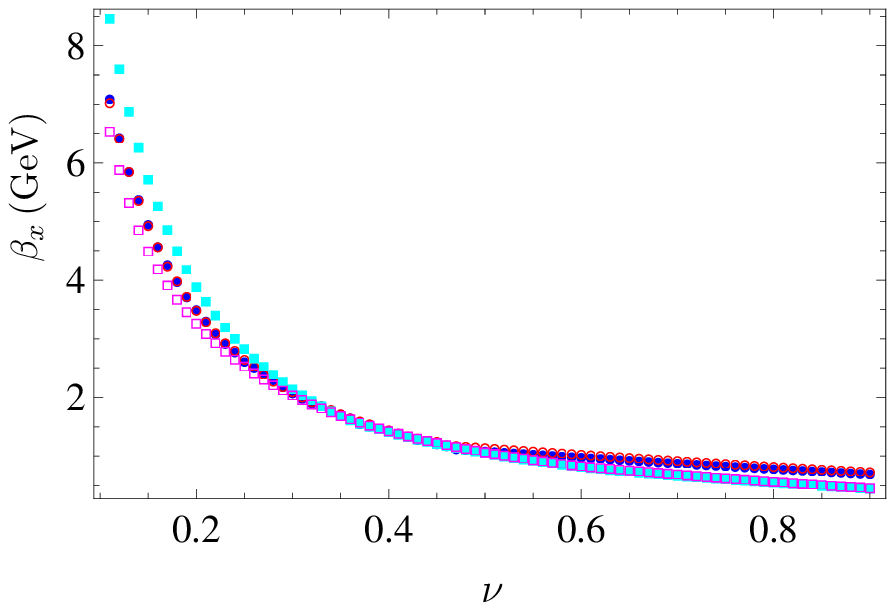}}
\caption{Comparison between the fitted quantities $\chi^2$, $c_1$, $c_0$ and $\beta_x$. The quantities which are obtained from the fitted orbital (the blue dots) and radial (the red circles) {\rt}  as the pion is included (five-point fit) are compared with the quantities obtained from the fitted orbital (the cyan solid squares) and radial (the magenta empty squares) {\rt}  as the $\pi^0$ is excluded (four-point fit). The used formulas are Eqs. (\ref{grform}) and (\ref{chi2}).}\label{fig:rthl}
\end{figure}

\subsection{Fit of the pion {\rts} by using five points}

In this subsection, the $\pi^0$ is included in fit, that is, five points or five states listed in Table \ref{tab:fitmass} are used to fit the orbital and radial {\rts} by employing Eq. (\ref{grform}), respectively.

The quantities $\chi^2$, $c_1$, $c_0$ and $\beta_{l}$ ($\beta_{n_r}$) vary with the exponent $\nu$, see Fig. \ref{fig:fit5}. $\nu$ ranges from $0.1$ to $0.9$ because as ${\nu}{\gtrapprox}1$, the fitted values of $c_0$ and $c_1$ become anomalous and omitting $\nu\in(0.9,1)$ does not affect the results.
$\chi^2$ increases with $\nu$. It increases rapidly and becomes large as $\nu{\gtrapprox}0.5$, see Fig. \ref{subfig:fiterr}.
$c_0$ is related with the curvature of the used formula (\ref{grform}). $c_0>0$ implies that the curvature of the formula (\ref{grform}) may be larger than expected while $c_0=0$ suggests that the exponent $\nu$ is appropriate or large. As shown in Fig. \ref{subfig:fitc0}, the fitted results prefer $0<\nu{\lessapprox}0.46$.
$c_1$ is expected to be greater than or equal to $0$, see the appendix \ref{subsec:str2} and Eq. (\ref{mrdef}). The $c_1-\nu$ plot shows that the better value is $\nu{\gtrapprox}0.43$, see \ref{subfig:fitc1}.
As $\nu\in(0.45,\,0.6)$, $\beta_l^2\in(1.436,\,0.942)$ and $\beta_{n_r}^2\in(1.54,\,1.04)$, see Fig. \ref{subfig:fitbeta}.
According to the previous discussions, for the five-point fit $\nu\approx0.45$ is the better fitted value without considering the Regge slopes.

\begin{figure}[!phtb]
\centering
\subfigure[]{\label{sf:fit4o}\includegraphics[scale=0.74]{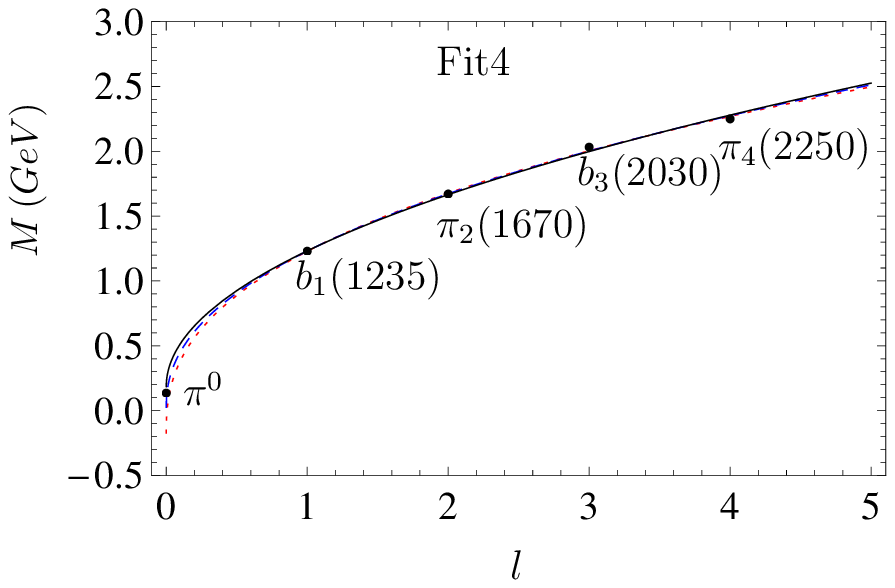}}
\subfigure[]{\label{sf:fit4r}\includegraphics[scale=0.7]{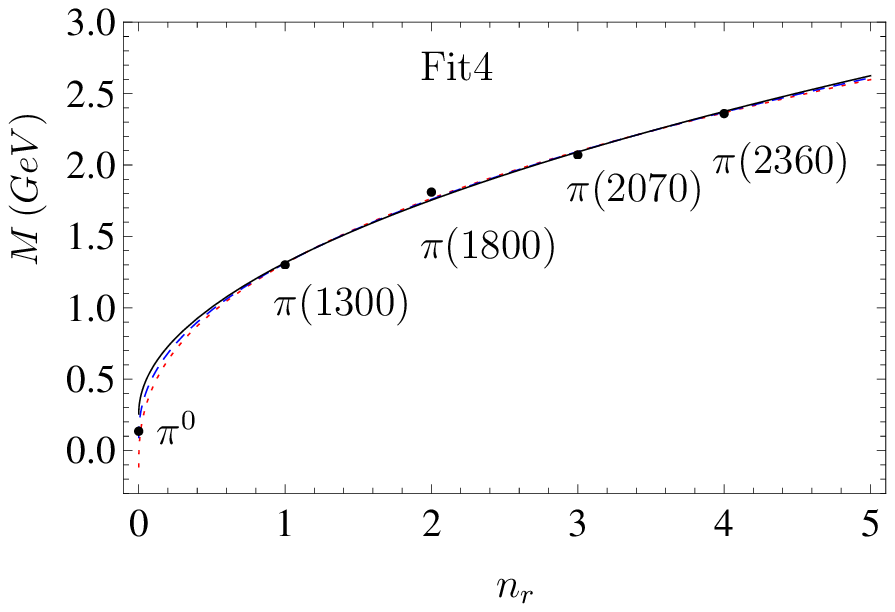}}
\subfigure[]{\label{sf:fit5o}\includegraphics[scale=0.7]{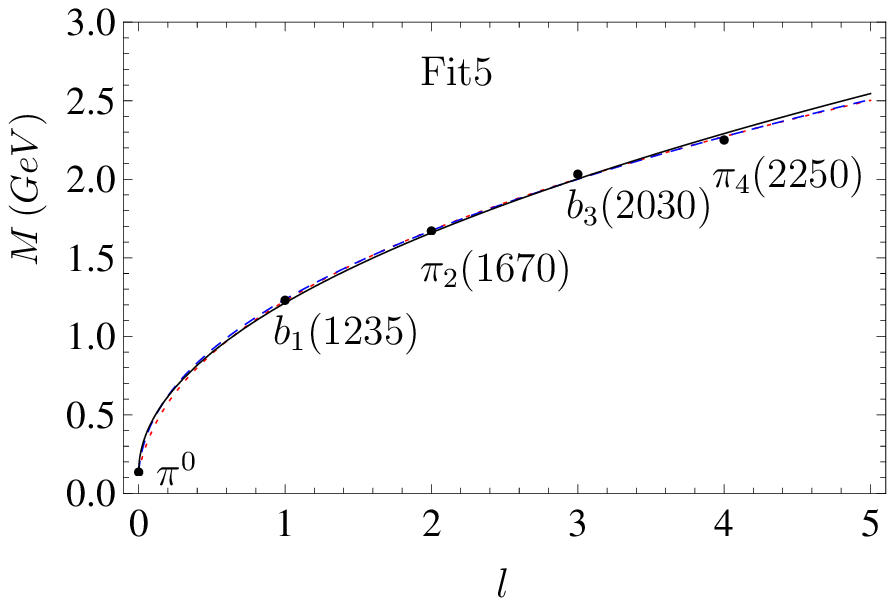}}
\subfigure[]{\label{sf:fit5r}\includegraphics[scale=0.7]{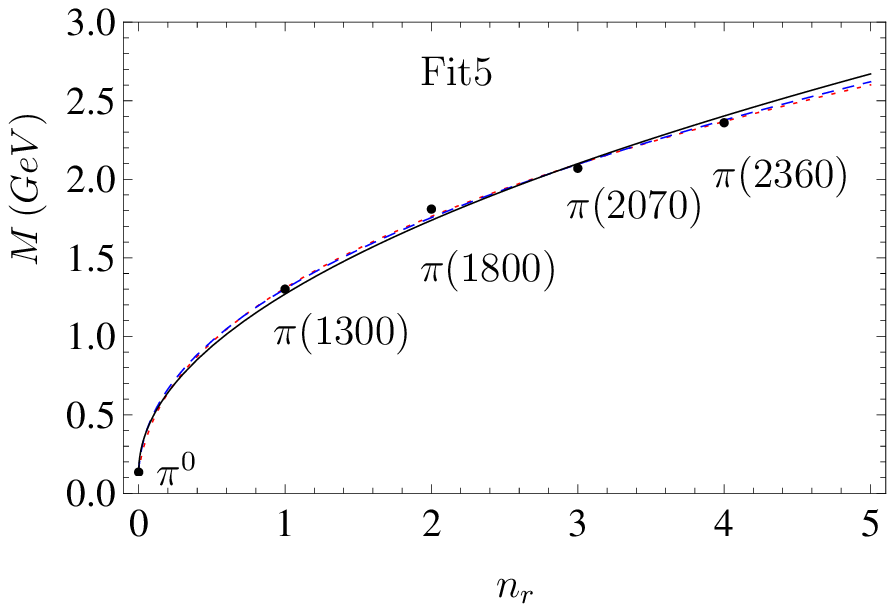}}
\caption{The fitted orbital and radial {\rts} for pion by using four points (Fit4) and by using five points (Fit5). They are plotted in the $(M,\,l)$ plane and in the $(M,\,n_r)$ plane, respectively. The red dotted lines are for $\nu=0.4$, the blue dashed lines are for $\nu=0.45$ and the black lines are for $\nu=0.5$. The used formulas are listed in Table \ref{tab:fitreg}.}\label{fig:regm}
\end{figure}

\begin{figure}[!phtb]
\centering
\subfigure[]{\label{sfs:fit4o}\includegraphics[scale=0.74]{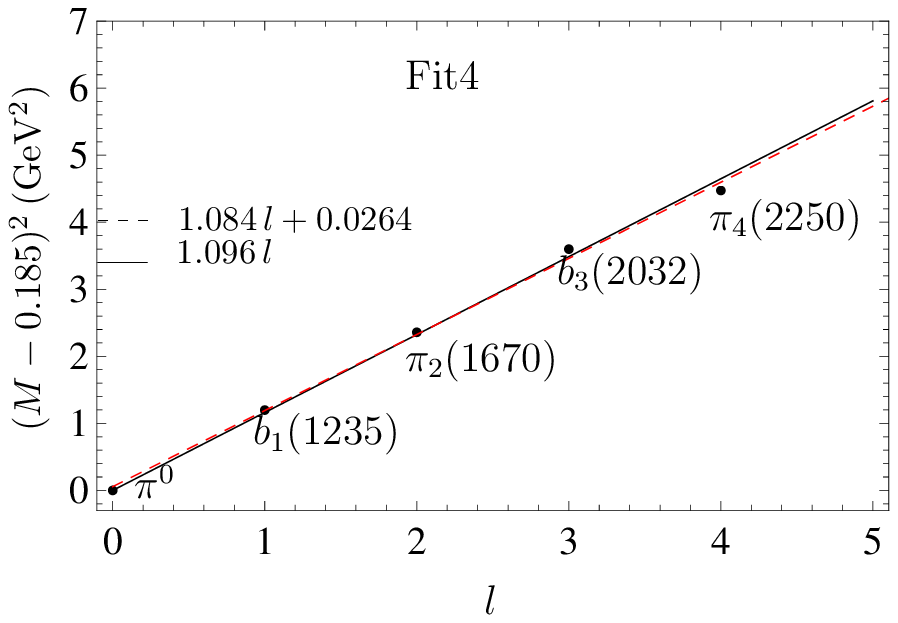}}
\subfigure[]{\label{sfs:fit4r}\includegraphics[scale=0.7]{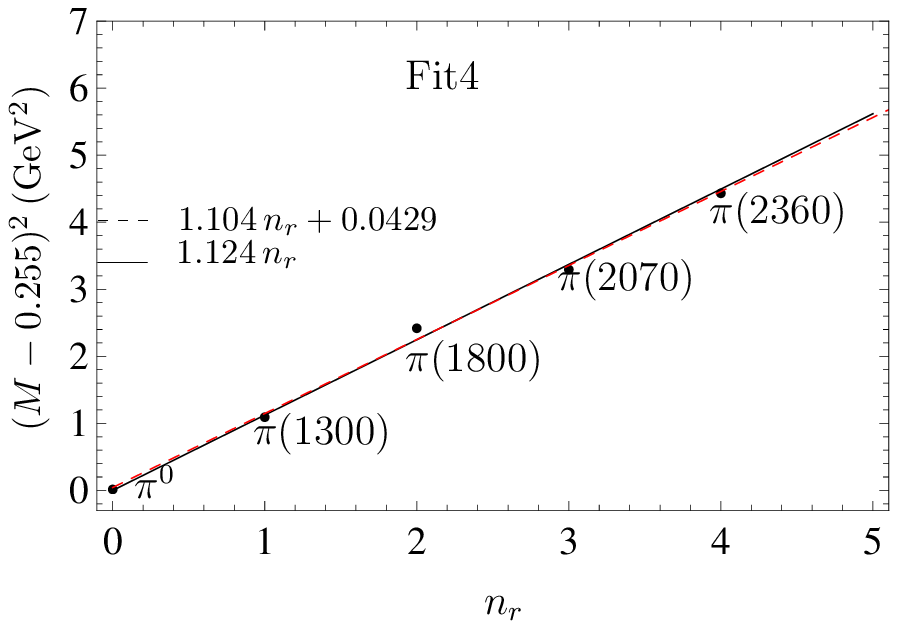}}
\subfigure[]{\label{sfs:fit5o}\includegraphics[scale=0.74]{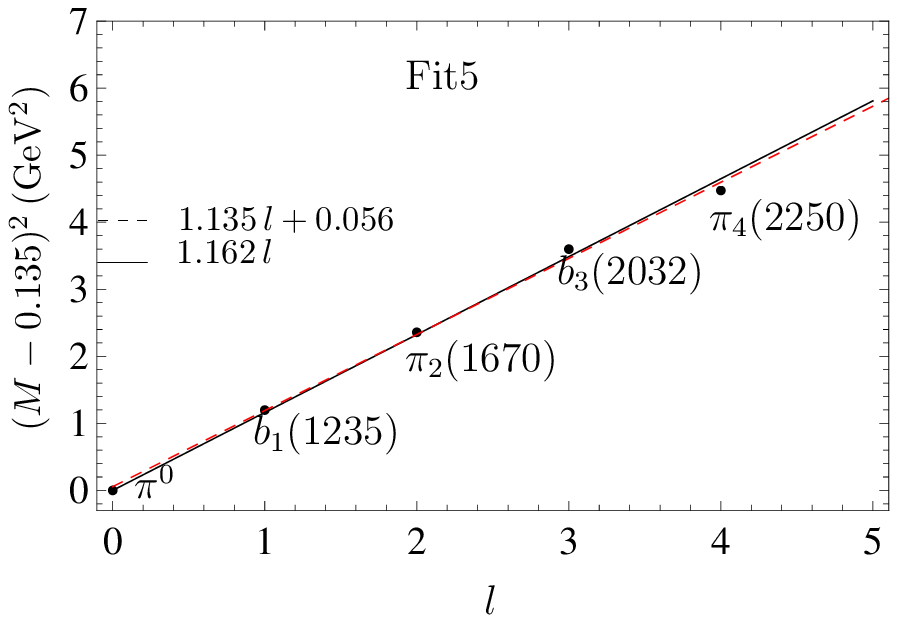}}
\subfigure[]{\label{sfs:fit5r}\includegraphics[scale=0.7]{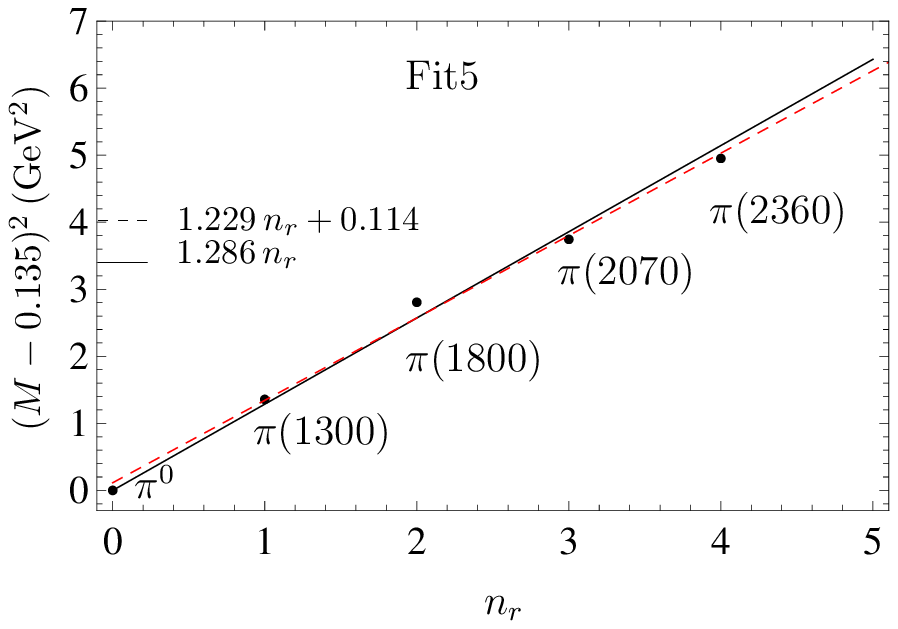}}
\caption{The fitted orbital and radial {\rts} for pion by using 4 points (Fit4) and by using 5 points (Fit5). They are plotted in the $((M-m_R)^2,\,l)$ plane and in the $((M-m_R)^2,\,n_r)$ plane, respectively. The used formulas for the black lines are from Table \ref{tab:fitreg}. The red dashed lines are the linear fits.}\label{fig:regm2}
\end{figure}

The fitted {\rts} by using five points are listed in Table \ref{tab:fitreg} with $\nu=0.4,\,0.45,\,0.5$. They are plotted in the $(M,\,l)$ plane and in the $(M,\,n_r)$ plane, respectively, see Fig. \ref{fig:regm}. The fitted masses are listed in Table \ref{tab:fitmass}, which are in agreement with the experimental values.

As mentioned in the introduction, the pion {\rts} are nonlinear in the $(M^2,\,x)$ ($x=l,\,n_r$) plane, see Fig. \ref{fig:regm2old}. However, both of the orbital pion {\rt} and the radial pion {\rt} can be described in a linear form when they are plotted in the $((M-m_R)^2,\,x)$ plane with appropriate $m_R$, see Fig. \ref{fig:regm2}.

\subsection{Fit of the pion {\rts} by using four points}
The pion can be explained as the pseudo-Nambu-Goldstone boson associated with chiral symmetry breaking \cite{Horn:2016rip,Sonnenschein:2018fph}. The low-mass pion is often excluded from the corresponding {\rts} formed by its orbitally or radially excited partners. In this subsection, we fit the orbital and radial {\rts} formed by the four orbitally excited states of $\pi^0$ and by the four radially excited states of $\pi^0$, respectively.

$\chi^2$ for the orbital and radial {\rts} increases with the exponent $\nu$. $\chi^2$ obtained by using the four points is roughly equal to $\chi^2$ calculated by using the five points as $\nu\in(0,0.5)$, see Fig. \ref{subfigure:fit4err}.
According to Fig. \ref{subfigure:fit4c0}, $\nu\in(0.39,\,0.64)$ for the orbital {\rt} and $\nu\in(0.15,\,0.6)$ for the radial {\rt} as $c_0=0$.
$c_1$ and $\beta_x$ calculated by four points have similar behavior as that obtained by using five points, respectively. In the range $\nu\in(0.3,0.5)$, $c_1$ (or $\beta_x$) obtained by using four points and five points are approximately equal.
As shown in Fig. \ref{subfigure:fit4c1}, $\nu{\gtrapprox}0.43$ for the radial {\rt} and $\nu{\gtrapprox}0.45$ for the orbital {\rt} as $c_1\ge0$.
As $\nu\in(0.45,\,0.6)$, $\beta_l^2\in(1.454,\,0.659)$ and $\beta_{n_r}^2\in(1.493,\,0.677)$, see Fig. \ref{subfigure:fit4beta}.
We can conclude that the better value of $\nu$ ranges from 0.45 to 0.5 for the four-point fit.

When $\pi^0$ is excluded in fit, the fitted {\rts} by using 4 points with $\nu=0.4,\,0.45,\,0.5$ agree well with the experimental values, see Table \ref{tab:fitreg}. The extrapolated masses of $\pi^0$ from the fitted radial {\rts} are $-115$ ${\rm MeV}$, $90$ ${\rm MeV}$ and $255$ ${\rm MeV}$ for $\nu=0.4,\,0.45,\,0.5$, respectively.
For the orbital {\rt}, the extrapolated masses of $\pi^0$ are $-175$ ${\rm MeV}$, $25$ ${\rm MeV}$ and $185$ ${\rm MeV}$ for $\nu=0.4,\,0.45,\,0.5$, respectively.
The fitted pion {\rts} are plotted in the $(M,\,l)$ plane and in the $(M,\,n_r)$ plane, respectively, see Fig. \ref{fig:regm}.
Whether the $\pi^0$ is included or not, both the orbital {\rt} and the radial {\rt} are linear when they are plotted in the $((M-m_R)^2,\,x)$ plane with nonzero $m_R$, see Fig. \ref{fig:regm2}. For the five-point fit, $m_R=135$ ${\rm MeV}$ for both the orbital {\rt} and the radial {\rt}, which is the experimental value of $\pi^0$. But for the four-point fit, $m_R=255$ ${\rm MeV}$ for the radial {\rt} and $m_R=185$ ${\rm MeV}$ for the orbital {\rt} in case of $\nu=0.5$.

\section{Discussions}\label{sec:dis}

\subsection{Confining potentials}
The confining potentials taking the power-law form with different power indexes are discussed in many works, such as $V_{conf}{\sim}r$ in the well-known Cornell potential \cite{Eichten:1974af}, $V_{conf}{\sim}r^{0.91}$ \cite{Flamm:1987kx}, $V_{conf}{\sim}r^{3/4}$ \cite{Lichtenberg:1988tn}, $V_{conf}{\sim}r^{2/3}$ \cite{Heikkila:1983wd,song:1991jg}, $V_{conf}{\sim}r^{1/2}$ \cite{Song:1986ix}, $V_{conf}{\sim}r^{0.1}$ \cite{Martin:1980jx,Martin:1980rm} and so on.
In Ref. \cite{Rai:2008sc}, the mass spectrums are extracted and the radial wave functions are reproduced from different models as well as from the {\nr} phenomenological quark antiquark potential of the type $V(r)=-\alpha_s/r+Ar^{\delta}$ with $\delta$ varying from 0.5 to 2. In Ref. \cite{Patel:2008na}, the power index range of $0.1<\delta<2.0$ has been explored  when computing the decay rates and spectroscopy of the $Q\bar{Q}$ mesons in the {\nr} potential.

The exponent $\nu$ in (\ref{grform}) is related to the confining potential, see Eq. (\ref{rtcom}) and the appendix \ref{subsec:pot}. In case of the {\ur} energy region, $\nu=1/2$ indicates the linear confining potential, $V_{conf}{\sim}r$. $\nu=0.45$ implies $V_{conf}{\sim}r^{9/11}$. $\nu=0.4$ arises from the confining potential $V_{conf}{\sim}r^{2/3}$.
In case of the {\nr} energy region, $\nu=1/2,\,0.45,\,0.4$ give $V_{conf}{\sim}r^a$ with $a=2/3,\,18/31,\,1/2$, respectively.
It is known that the orbitally excited states and the radially excited states of pion are taken as the {\ur} systems \cite{Chen:2021kfw}. The suggested $0.45\le\nu\le0.5$ by the fit indicates the confining potential $r^a$ with $9/11{\le}a\le1$.

\subsection{Parameter $c_1$ or $m_R$}

In the {\ur} limit, $c_1$ in (\ref{grform}) or $m_R$ in (\ref{reglike}) is usually assumed to be zero, i.e., the {\rt} takes the form $M^2={\alpha_x}(x+c_0)^{\gamma}$ with $\gamma=1$. According to Eqs. (\ref{sse}), (\ref{massform}), (\ref{reglike}), (\ref{mrdef}), (\ref{lowMass}) and (\ref{highMass}), $c_1$ or $m_R$ is related with the masses of constituents and the constant part of the interaction energy, especially in the {\nr} energy region. In Ref. \cite{Godfrey:1985xj}, $m_u=m_d=220$ ${\rm MeV}$, $C$ in Eq.
(\ref{potv}) reads $C=-253$ ${\rm MeV}$. Substituting $m_u$, $m_d$ and $\epsilon_c=C$ into Eq. (\ref{mrdef}) gives $c_1=m_R=187$ ${\rm MeV}$. It is in excellent agreement with $185$ ${\rm MeV}$ which is obtained from the fitted orbital {\rts} ($\nu=0.5$) and is smaller than $255$ ${\rm MeV}$ which is from the fitted radial {\rts} ($\nu=0.5$) as $\pi^0$ is excluded in fit, see Table \ref{tab:fitreg}.

As shown in the appendix \ref{appsec:b}, nonzero $m_R$, i.e., nonzero $c_1$ in Eq. (\ref{grform}) will shift the slope of the {\rts} to a lower value. They can give the reasonable slopes. It shows that nonzero $c_1$ or nonzero $m_R$ is appropriate and acceptable.

It is the nonzero $c_1$ or $m_R$ together with $\beta_x(x+c_0)^{1/2}$ that leads to the nonlinearity of the orbital $(M^2,\,l)$ pion {\rt} and the nonlinearity of the radial $(M^2,\,n_r)$ pion {\rt}.
As $m_R$ is not equal to zero and is chosen appropriately, the radial pion {\rt} in the $((M-m_R)^2,\,n_r)$ plane and the orbital {\rt} in the $((M-m_R)^2,\,l)$ plane are linear whether the $\pi^0$ is included on the {\rts}, see Fig. \ref{fig:regm2}.

According to Eq. (\ref{grform}) or (\ref{reglike}), $c_1{\ne}0$ [or $m_R{\ne}0$] means that one part of $M$ keeps constant and does not vary with $l$ and $n_r$ while the other portion changes with $l$ or $n_r$. $c_1=0$ [or $m_R=0$] implies that all parts of the bound-state masses are effected by the potentials because $\nu$ is related with the confining potential.
$c_0=0$ indicates that $M$ varies with $l$ or $n_r$ in a simple way. As $c_0{\ne}0$, $c_0$ will be entangled with $l$ ($n_r$) because $\nu{\ne}1$.

\subsection{A note on $\pi^0$}
There are lots of discussions on the intrinsic structure of the pion \cite{Horn:2016rip,Arrington:2021biu,Fariborz:2021gtc}.
%
%
In Ref. \cite{Alexandrou:2017itd}, the lattice QCD gives $m_{\pi}=296$ ${\rm MeV}$. In Ref. \cite{Santowsky:2021ugd}, $m_{\pi}=328$ ${\rm MeV}$. In Ref. \cite{Frezzotti:2022dwn}, the pion mass is in the range $250-500$ ${\rm MeV}$. The extrapolated mass are $m_{\pi}=185$ ${\rm MeV}$ for the orbital {\rt} by using four points and
$m_{\pi}=255$ ${\rm MeV}$ for the radial {\rt} by using four points in case of the linear confining potential. They are larger than the experimental result and smaller than the results in Refs. \cite{Alexandrou:2017itd, Santowsky:2021ugd}. According to the discussions in section \ref{sec:fit}, it is not foreclosed and reasonable that $\pi^0$ is taken as the first point on the pion {\rts}, see Figs. \ref{sf:fit4o}, \ref{sf:fit4r}, \ref{sfs:fit4o} and \ref{sfs:fit4r}. It implies that $\pi^0$ can be regarded as the quark-antiquark state like other states on the pion {\rts}.

\section{Conclusions}\label{sec:conclusions}
The orbital and radial pion {\rts} are fitted phenomenologically by employing the ansatz $M={\beta_x}\left(x+c_0\right)^{\nu}+c_1$ where $x=l,\,n_r$.
It is shown that nonzero $m_R$ is reasonable and acceptable. Nonzero $m_R$ or $c_1$ gives an explanation that the pion {\rts} are concave in the usually employed $(M^2,\,x)$ plane as being examined more precisely. As $m_R$ is chosen appropriately, both the orbital and radial pion {\rts} are linear in the $((M-m_R)^2,\,x)$ plane whether the $\pi^0$ is included or not on the {\rts}.
It is reasonable and not foreclosed that $\pi^0$ is regarded as the first point on the pion {\rts}.
The fitted pion {\rts} suggest $0.45\le\nu\le0.5$. It indicates the confining potential $r^a$ with $9/11{\le}a\le1$.

We present discussions in the appendix \ref{appenda:structa} on the structure of the {\rts} plotted in the $(M,\,x)$ plane and in the $((M-m_R)^2,\,x)$ plane based on the potential models and the string models.
In the appendix \ref{appsec:b}, the {\rts} for the light nonstrange mesons with different $m_R$ are shown in the $((M-m_R)^2,\,x)$ plane.
It is illustrated that $m_R$ can be nonzero for the light nonstrange mesons.

\section*{Acknowledgments}
We are very grateful to the anonymous referees for the valuable comments and suggestions. This work is supported by the Natural Science Foundation of Shanxi Province of China under Grant no. 201901D111289, which is sponsored by the Shanxi Science and Technology Department.

\appendix

\section{Structure of the {\rts}}\label{appenda:structa}

The potential models are the basic tools of the phenomenological approach to model the features of QCD relevant to hadron with the aim to produce concrete results. In Ref. \cite{Chen:2021kfw}, we present discussions on the structure of the meson {\rts} plotted in the $(M^2,\,x)$ plane where $x=n_r,\,l$ based on the quadratic form of the spinless Salpeter-type equation \cite{Baldicchi:2007ic,Baldicchi:2007zn,Brambilla:1995bm,chenvp,chenrm}. Herein, we present discussions on the structure of the meson {\rts} plotted in the $(M,\,x)$ plane and in the $((M-m_R)^2,\,x)$ plane \cite{note}.

\subsection{Structure of the {\rts} in the $(M,\,x)$ plane}

\subsubsection{Potential models}\label{subsec:pot}

The relativistic quark model or the Godfrey-Isgur (GI) model is employed. The spin-dependent interactions are not considered. The dynamic equation is the {\sse} (SSE) \cite{Durand:1981my,Durand:1983bg,Lichtenberg:1982jp,Godfrey:1985xj,Jacobs:1986gv} which reads
\begin{eqnarray}\label{sse}
M\Psi({\bfr})=\left(\omega_1+\omega_2\right)\Psi({\bfr})+V\Psi({\bfr}),
\end{eqnarray}
where $M$ is the bound state mass, $\omega_i$ is the square-root operator of the relativistic kinetic energy of constituent
\bea\label{omega}
\omega_i=\sqrt{m_i^2+{\bf p}^2}=\sqrt{m_i^2-\Delta},
\eea
$m_1$ and $m_2$ are the effective masses of the constituents, respectively.
In the present work, the Cornell potential \cite{Eichten:1974af} is considered,
\bea\label{potv}
V(r)=-\frac{\alpha}{r}+{\sigma}r+C,
\eea
where $\sigma$ is the string tension. $\alpha=4\alpha_s/3$ and $\alpha_s$ is the strong coupling constant of the color Coulomb potential. $C$ is a parameter which is fundamental and indispensable as the quark masses, slope of the linear potential $\sigma$, and the strong coupling
constant. $C{\approx}-2\sqrt{\sigma}\exp[-\gamma_E+1/2]$ \cite{Gromes:1981cb,Lucha:1991vn} where $\gamma_E$ is the Euler constant.

In the {\nr} (NR) region, $m_1,m_2{\gg}{\bfpa}$, we can obtain from Eq. (\ref{sse})
\begin{eqnarray}\label{ssenr}
M\Psi({\bfr})=\left(m_1+m_2+\frac{{\bf p}^2}{2\mu}\right)\Psi({\bfr})+V\Psi({\bfr}),
\end{eqnarray}
where $\mu=m_1m_2/(m_1+m_2)$. By employing the Bohr-Sommerfeld quantization approach \cite{Brau:2000st,brsom}, we obtain from Eq. (\ref{ssenr})
\begin{align}\label{rghh}
M{\sim}&\frac{3}{2}\left(\frac{\sigma^2}{\mu}\right)^{1/3}l^{2/3}\quad (l{\gg}n_r),\nonumber\\
M{\sim}&\left(\frac{3\pi}{2}\right)^{2/3}\left(\frac{\sigma^2}{2\mu}\right)^{1/3}n_r^{2/3}
\quad (n_r{\gg}l).
\end{align}
In the {\ur} (UR) region, ${\bfpa}{\gg}m_1,m_2$, we can obtain from Eq. (\ref{sse})
\begin{eqnarray}\label{sseur}
M\Psi({\bfr})=2{\bfpa}\Psi({\bfr})+V\Psi({\bfr}).
\end{eqnarray}
Then we have from Eq. (\ref{sseur})
\begin{align}\label{rgll}
M{\sim}&2\sqrt{2\sigma}\sqrt{l}\quad (l{\gg}n_r),\nonumber\\
M{\sim}&2\sqrt{{\pi}\sigma}\sqrt{n_r}\quad (n_r{\gg}l).
\end{align}
Both of Eqs. (\ref{rghh}) and (\ref{rgll}) have been obtained in Ref. \cite{Brau:2000st}.

If one or both of the constituents are in the intermediate (IM) energy region, $m_i{\sim}{\bfpa}$ or $m_1,m_2{\sim}{\bfpa}$. According to the author's knowledge, the approximated form of $M$ has not been obtained due to its complexity. If there is a simple approximation
\bea\label{expinr}
M{\sim}x^{\nu}\quad (x=l,\,n_r),
\eea
$\nu$ is expected to lie between $1/2$ and $2/3$ \cite{MartinContreras:2020cyg}.

\begin{figure}[!phtb]
\centering
\includegraphics[scale=0.9]{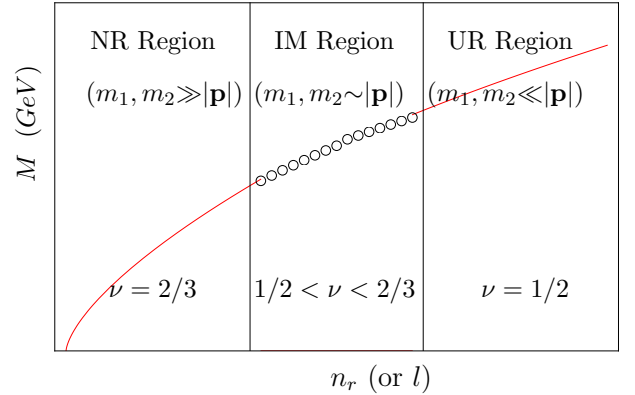}
\caption{The structure of the {\rt} corresponding to different energy regions according to Eqs. (\ref{grform}) and (\ref{rtcomb}). It is plotted in the $(M,\,x)$ plane where $x=n_r,\,l$. The confining potential is assumed to be linear ${\sigma}r$. The NR region represents the {\nr} region, the IM region denotes the intermediate region and the UR region is the {\ur} region. The circles represent the part of the {\rt} in the intermediate region which remains unclear.}\label{fig:struct}
\end{figure}

Based on Eqs. (\ref{ssenr}), (\ref{rghh}), (\ref{sseur}), (\ref{rgll}) and (\ref{expinr}),  we can propose a generic form of a {\rt} which has the same form as the new ansatz in Eq. (\ref{grform}). If the confining potential is linear, $V_{conf}={\sigma}r$, the theoretical values of the exponent $\nu$ read
\begin{eqnarray}\label{rtcomb}
\left\{\begin{array}{cc}
\nu=\frac{2}{3}, & \text{NR region}, \\
\frac{1}{2}<\nu<\frac{2}{3}, & \text{IM region},\\
\nu=\frac{1}{2}, & \text{UR region}.
\end{array}\right.
\end{eqnarray}
The plot corresponding to Eq. (\ref{rtcomb}) is shown in Fig. \ref{fig:struct}.
If the confining potential is the power-law potential
\bea
V_{conf}={\sigma}r^a,
\eea
Eq. (\ref{rtcomb}) becomes \cite{Brau:2000st}
\begin{eqnarray}\label{rtcom}
\left\{\begin{array}{cc}
\nu=\frac{2a}{a+2}, & \text{NR region}, \\
\frac{a}{a+1}<\nu<\frac{2a}{a+2}, & \text{IM region},\\
\nu=\frac{a}{a+1}, & \text{UR region}.
\end{array}\right.
\end{eqnarray}
Different forms of kinematic terms corresponding to different energy regions will yield different behaviors of the {\rts} \cite{Chen:2021kfw}. ${\bf p}$ and $r^a$ leads to $M{\sim}x^{a/(a+1)}$ while ${\bf p}^2$ and $r^a$ gives $M{\sim}x^{2a/(a+2)}$ $(x=l,\,n_r)$.

The behaviors of the {\rts} obtained from the SSE of the GI model [Eq. (\ref{sse})], Eqs. (\ref{rtcomb}) and (\ref{rtcom}), are consistent with the results obtained in Refs. \cite{Martin:1986rtr,Lucha:1991vn} and are consistent with the results from other potential models, such as the quadratic form of the spinless Salpeter-type equation \cite{Chen:2021kfw,Chen:2018hnx,Chen:2018bbr,Chen:2018nnr}, the Schr\"{o}dinger equation \cite{Brau:2000st,FabreDeLaRipelle:1988zr,Quigg:1979vr,Hall:1984wk}, the Dirac equation \cite{Olsson:1994cv}, the Klein-Gordon equation \cite{Kang:1975cq,Sharma:1982ez,Kulikov:2006gg}, the relativistic Thompson equation \cite{Kahana:1993yd}, a first principle Salpeter equation \cite{Baldicchi:1998gt,Baldicchi:1999vr}, a three-dimensional reduction of the Bethe-Salpeter equation \cite{DiSalvo:1994mf} and so on.

\subsubsection{String Models}
For comparison, we list in this subsection the results obtained in Ref. \cite{Sonnenschein:2018fph}. Based on the holography inspired stringy hadron model, the following equations are derived from the relation between angular momentum and energy
\begin{align}
 M=& \sum_{i=1,2}\left(\frac{m_i}{\sqrt{1-v_i^2}} + T\ell_i\frac{\arcsin{v_i}}{v_i}\right)\,, \nonumber\\
J + n -a  =& \sum_{i=1,2}\left(\frac{m_i v_i \ell_i}{\sqrt{1-v_i^2}}\right.
\end{align}
\begin{align}
 &\left.+ \frac12T\ell_i^2\frac{\arcsin{v_i}-v_i\sqrt{1-v_i^2}}{v_i^2}\right), \label{eq:MJ}
\end{align}
where \(T\) is the string tension, \(v_i\) is the velocity of the endpoint with the mass \(m_i\), and \(\ell_i\) is the distance of the mass from the center of mass around which the endpoint particles rotate. \(v_i\) are related to each other,
\bea
\omega = \frac{v_1}{\ell_1} = \frac{v_2}{\ell_2}\,,
\eea
and the boundary conditions of the string imply
\bea \frac{T\ell_i}{m_i} = \frac{v_i^2}{1-v_i^2}\,.
\eea

In the high energy limit, \(v \rightarrow 1\). The authors \cite{Sonnenschein:2018fph} give an expansion in $m/M$ in the symmetric case \(m_1=m_2=m\),
\begin{align}
 J + n - a  &= \alp M^2\left[1-\frac{8\sqrt{\pi}}{3}\left(\frac{m}{M}\right)^{3/2} \right.\nonumber\\
 &\left.+ \frac{2\sqrt{\pi^3}}{5}\left(\frac{m}{M}\right)^{5/2} + \cdots\right], \label{lowMass}
\end{align}
where \(\alp = (2\pi T)^{-1}\). The opposing low energy limit, \(v \rightarrow 0\), holds when \((M-2m)/2m \ll 1\). The expansion is \cite{Sonnenschein:2018fph,Kruczenski:2004me}
\begin{align}
 J + n - a  &= \frac{4\pi}{3\sqrt{3}}\alp m^{1/2} (M-2m)^{3/2}\nonumber\\
  &+ \frac{7\pi}{54\sqrt{3}} \alp m^{-1/2} (M-2m)^{5/2} + \cdots \label{highMass}
\end{align}

The {\rts} obtained from the potential models, see Eq. (\ref{rtcomb}), are consistent with the results obtained from the holography inspired stringy hadron model \cite{Sonnenschein:2018fph} [see Eqs. (\ref{lowMass}) and (\ref{highMass})], the Holographic dual of large-$N_c$ QCD \cite{Kruczenski:2004me}, the relativistic flux tube model \cite{Cotugno:2009ys,Burns:2010qq,Selem:2006nd}, the Nambu string model \cite{Nambu:1974zg}, the string-like model \cite{Afonin:2014nya} and so on.
They are also in agreement with other models, such as the holographic Ads/QCD context \cite{MartinContreras:2020cyg,Karch:2006pv}, the light-front holographic QCD \cite{Brodsky:2016yod}, the holographic model within deformed AdS$_5$ space metrics \cite{FolcoCapossoli:2019imm} and so on.

\subsection{Structure of the {\rts} in the $((M-m_R)^2,\,x)$ plane}
\label{subsec:str2}

The mass of a meson can be written as
\bea\label{massform}
M=m_1+m_2+\epsilon,
\eea
where $\epsilon$ is the interaction energy. Suppose $\epsilon$ can be divided into a constant $\epsilon_c$ and a nonconstant function $\epsilon_f$,
\bea
\epsilon=\epsilon_c+\epsilon_f.
\eea
Subtract $m^{\prime}_R$ on both sides of Eq. (\ref{massform}), then square both sides of the obtained equation. This gives
\bea\label{form1a}
(M-m^{\prime}_R)^2=\delta^2+2{\delta}\epsilon_f+\epsilon_f^2,\quad \delta=m_1+m_2+\epsilon_c-m^{\prime}_R.
\eea
If $|\delta|{\sim}\epsilon_f$, none of the three terms on the right side of Eq. (\ref{form1a}) can be omitted, and there is
\bea\label{fredo}
(M-m^{\prime}_R)^2{\sim}\epsilon_f^2,\,\epsilon_f.
\eea
If $m^{\prime}_R$ makes $|\delta|{\gg}\epsilon_f$, $\delta^2$ is dominant and $\epsilon_f^2$ can be neglected, then (\ref{form1a}) becomes
\bea\label{form1b}
(M-m^{\prime}_R)^2=\delta^2+2{\delta}\epsilon_f,
\eea
and there is
\bea\label{fredef}
(M-m^{\prime}_R)^2{\sim}\epsilon_f.
\eea
If $m^{\prime}_R$ makes $|\delta|{\ll}\epsilon_f$ or $\delta=0$, $\epsilon_f^2$ plays dominant role while $\delta^2$ and $2{\delta}\epsilon_f$ can be neglected, then (\ref{form1a}) becomes
\bea\label{form1c}
(M-m^{\prime}_R)^2=\epsilon_f^2,
\eea
and there is
\bea\label{rgredft}
(M-m^{\prime}_R)^2{\sim}\epsilon_f^2.
\eea
If $m^{\prime}_R=0$, (\ref{form1a}) becomes the conventional form of the {\rts}, $M^2=f(l,\,n_r)$. [And the structure of the {\rts} in the form $M^2=f(l,\,n_r)$ has been discussed in Ref. \cite{Chen:2021kfw}.]
According to Eqs. (\ref{fredo}), (\ref{fredef}) and (\ref{rgredft}), different choices of $m^{\prime}_R$ result in different behaviors.
The necessary cautions should be taken in using the formula (\ref{reglike}) to fit a {\rt}. It is suggested that using the formula (\ref{grform}) to fit the {\rts} and then transforming the fitted results into the form in (\ref{reglike}).

\begin{figure}[!phtb]
\centering
\includegraphics[scale=0.9]{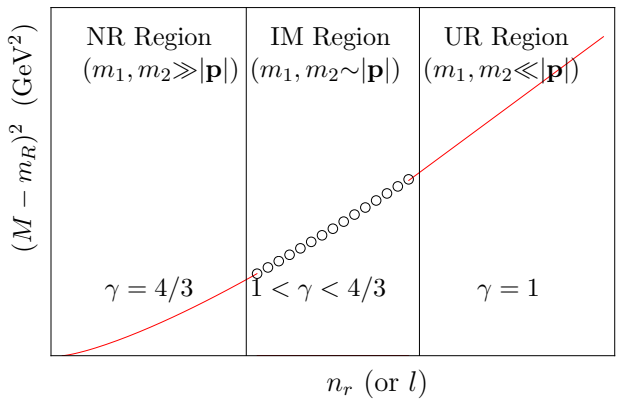}
\caption{The structure of the {\rt} in the $((M-m_R)^2,\,x)$ plane where $x=n_r,\,l$. The behavior in different energy regions accords to Eqs. (\ref{reglike}) and (\ref{rglsimb}). The confining potential is assumed to be linear ${\sigma}r$. The circles represent the part of the {\rt} in the intermediate region which remains unclear.}\label{fig:likestruct}
\end{figure}

Eq. (\ref{form1c}) will lead to a generic form of the Regge-like formula which reads \cite{Chen:2021kfw}
\bea\label{reglike}
(M-m_R)^2=\alpha_x(x+c_0)^{\gamma}\quad (x=l,\,n_r),
\eea
where
\bea\label{mrdef}
m_R=m_1+m_2+\epsilon_c.
\eea
Eq. (\ref{reglike}) is an extension of the Regge-like formulas in Refs. \cite{Veseli:1996gy,Afonin:2014nya,Afonin:2020bqc,Chen:2017fcs,Jia:2019bkr,Jia:2018yvk,Chen:2014nyo,Afonin:2013hla}.
Using Eqs. (\ref{rtcom}) and (\ref{reglike}), we have for the power-law potentials
\begin{eqnarray}\label{rglsim}
\left\{\begin{array}{cc}
\gamma=\frac{4a}{a+2}, & \text{NR region}, \\
\frac{2a}{a+1}<\gamma<\frac{4a}{a+2}, & \text{IM region},\\
\gamma=\frac{2a}{a+1}, & \text{UR region}.
\end{array}\right.
\end{eqnarray}
For the linear confining potential, Eq. (\ref{rglsim}) becomes
\begin{eqnarray}\label{rglsimb}
\left\{\begin{array}{cc}
\gamma=\frac{4}{3}, & \text{NR region}, \\
1<\gamma<\frac{4}{3}, & \text{IM region},\\
\gamma=1, & \text{UR region}.
\end{array}\right.
\end{eqnarray}
It is shown in Fig. \ref{fig:likestruct}.

Eq. (\ref{reglike}) can alao be obtained from the new ansatz in Eq. (\ref{grform}), where $m_R=c_1$, $\alpha_x=\beta^2_x$, $\gamma=2\nu$. In addition, with the help of the Taylor series, the new ansatz $M = \beta_x(x + c_0)^{\nu} +c_1$ in Eq. (\ref{grform}) can be approximated as the form of $(M-c_1)^2 \approx 2\nu\beta_x^2c_0^{2\nu-1}x+ \beta_x^2 c_0^{2\nu}$ type Regge trajectory when $\nu \ne 0.5$ if $c_0$ is large and the approximation becomes equal when $\nu=0.5$. Similarly, the new ansatz $M = \beta_x(x + c_0)^{\nu} + c_1$ can be approximated as the conventional form of Regge trajectory $M^2 \approx 2\nu\beta_xc_0^{\nu-1}(\beta_xc_0^\nu+c_1)x + (\beta_xc_0^{\nu}+c_1)^2$ ($\nu>0$). If $\beta_x(x + c_0)^{\nu}{\ll}c_1$, there is $M^2\approx 2c_1\beta_x(x+c_0)^{\nu}+c^2_1$ \cite{Chen:2021kfw}. If $\beta_x(x + c_0)^{\nu}{\gg}c_1$, there is $M^2\approx \beta_x^2(x+c_0)^{2\nu}$.

The new ansatz in Eq. (\ref{grform}) can be rewritten in a more general form
\bea\label{mgrform}
M=\left({\alpha'_l}\,l+{\alpha'_{n_r}}n_r +c'_0\right)^{\nu}+c_1.
\eea
Correspondingly, Eq. (\ref{reglike}) has the general form
\bea\label{reglikeg}
(M-m_R)^2=({\alpha'_l}\,l+{\alpha'_{n_r}}n_r +c'_0)^{\gamma},
\eea
which can be obtained from Eq. (\ref{mgrform}). When $l\ne0$ and $n_r\ne0$ simultaneously exist, Eqs. (\ref{mgrform}) and (\ref{reglikeg}) work evidently. As expected, ${\alpha'_l}\,l+c'_0$ in the {\rts} increases with $l$ and ${\alpha'_{n_r}}n_r +c'_0$ increases with $n_r$, see Figs. \ref{fig:reggemr2}, \ref{fig:reggemo2} and Table \ref{tab:regmr2}.

\section{$((M-m_R)^2,\,x)$ {\rts} for the light nonstrange mesons}\label{appsec:b}

\begin{figure*}[!phtb]
\centering
\subfigure[]{\label{subfigure:rpi0}\includegraphics[scale=0.54]{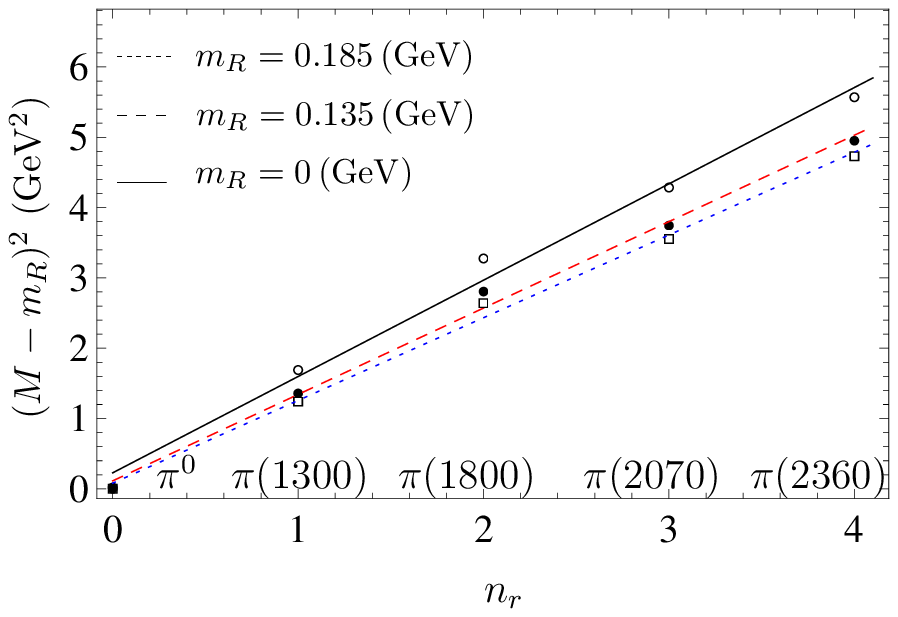}}
\subfigure[]{\label{subfigure:rrho7}\includegraphics[scale=0.54]{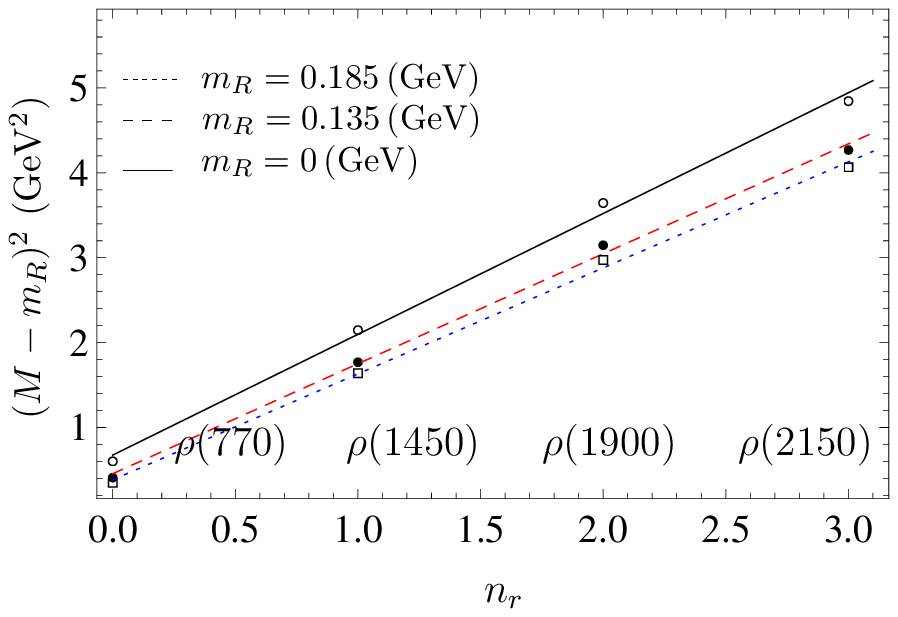}}
\subfigure[]{\label{subfigure:reta}\includegraphics[scale=0.54]{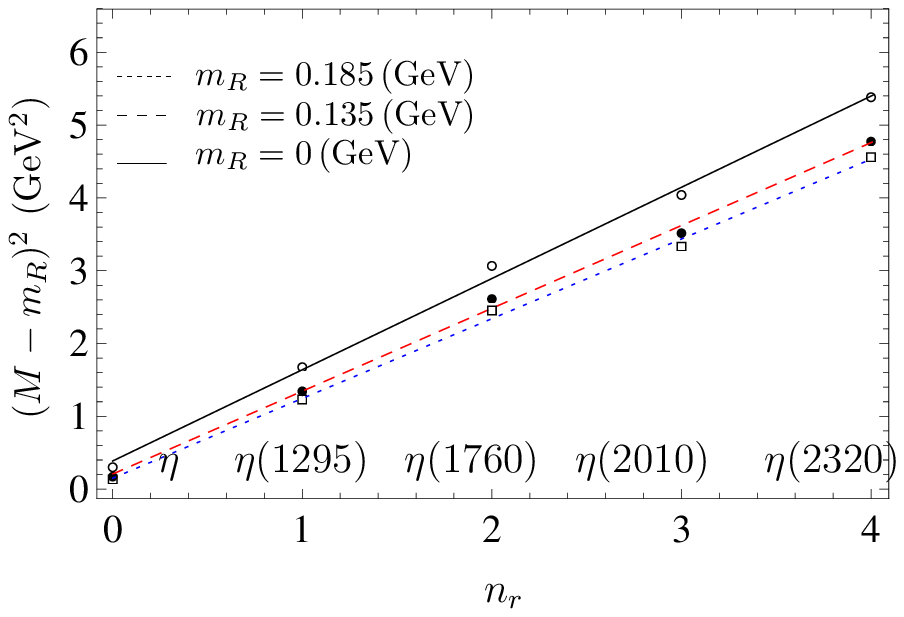}}
\subfigure[]{\label{subfigure:romega}\includegraphics[scale=0.54]{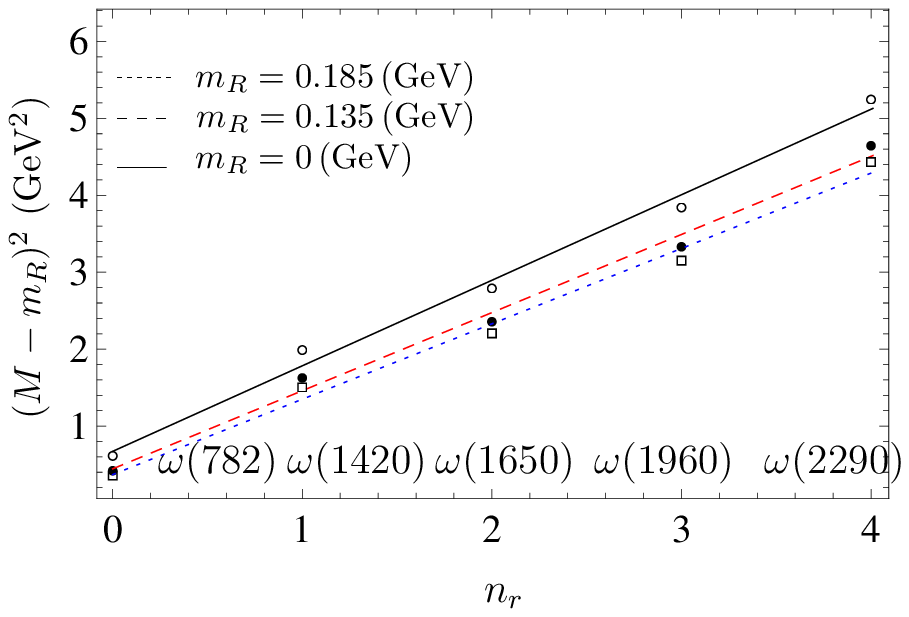}}
\subfigure[]{\label{subfigure:ra1}\includegraphics[scale=0.54]{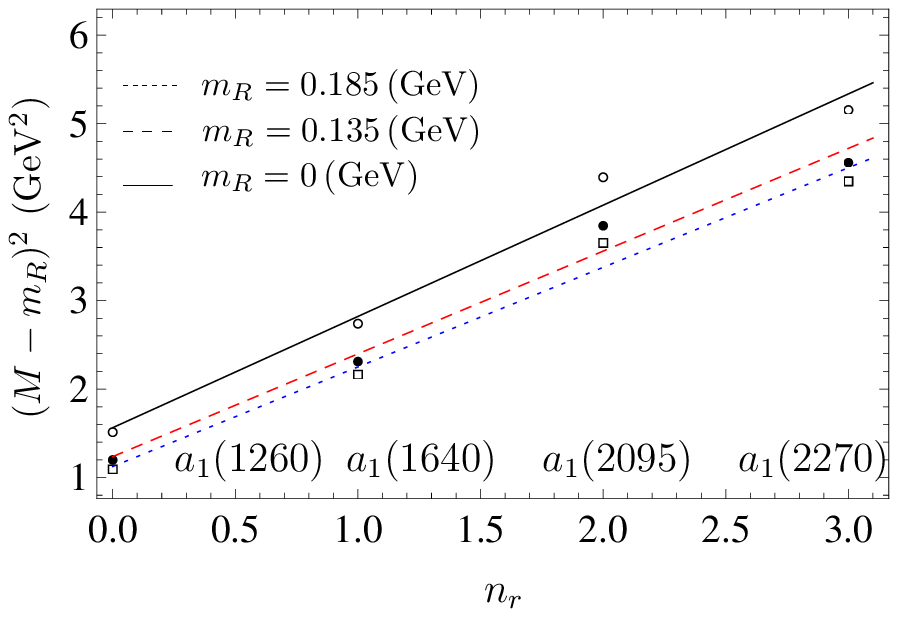}}
\subfigure[]{\label{subfigure:ra2}\includegraphics[scale=0.54]{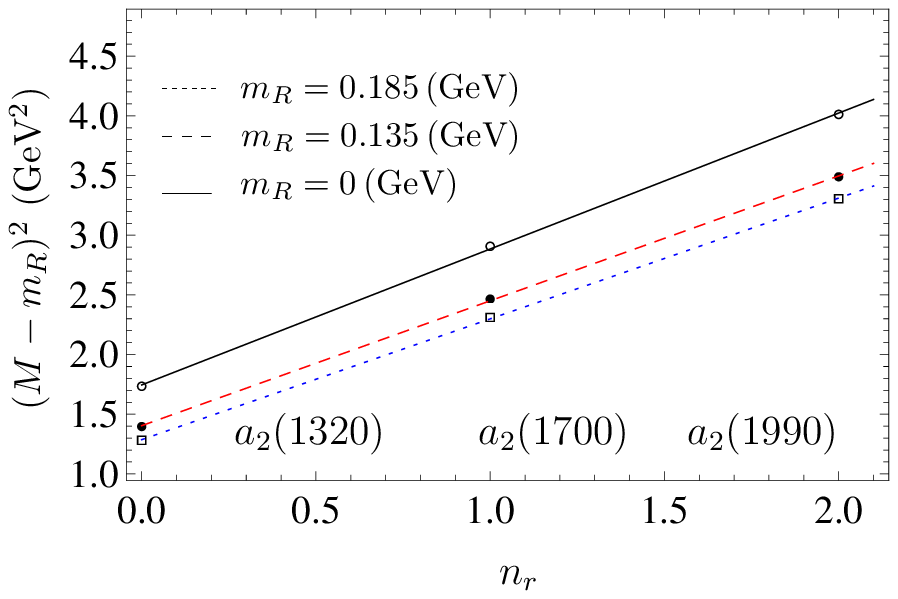}}
\subfigure[]{\label{subfigure:rb1}\includegraphics[scale=0.54]{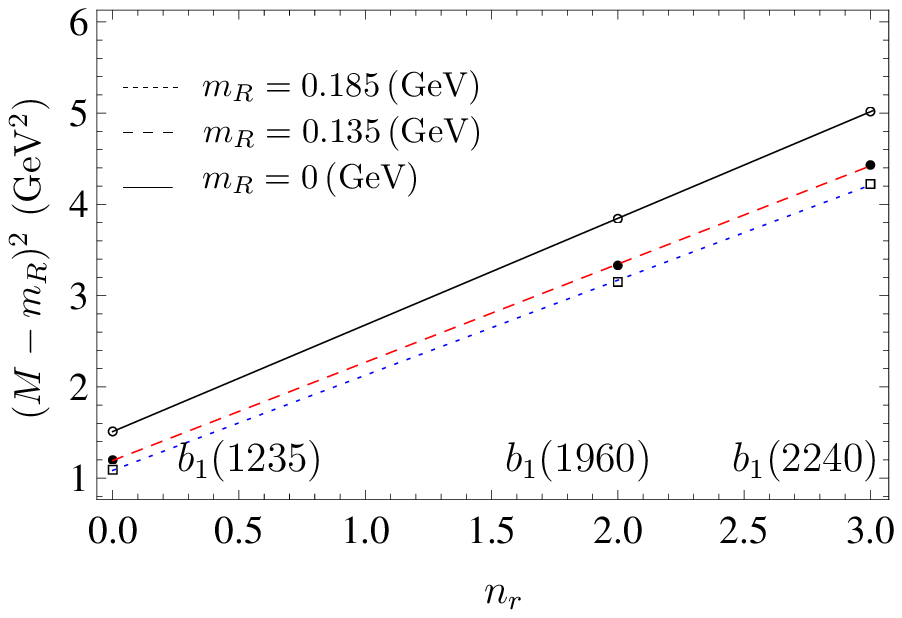}}
\subfigure[]{\label{subfigure:rf2}\includegraphics[scale=0.54]{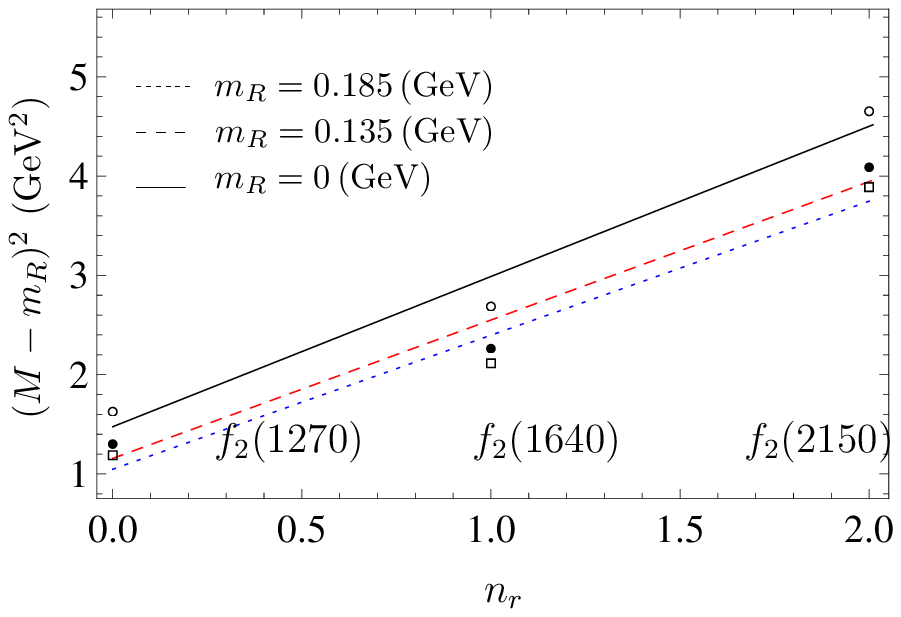}}
\subfigure[]{\label{subfigure:rh1}\includegraphics[scale=0.54]{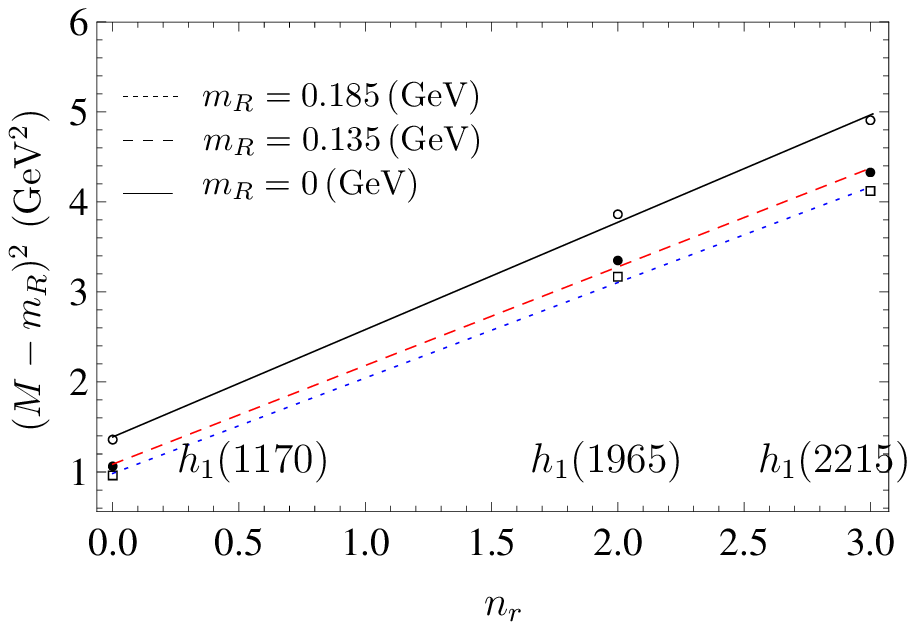}}
\subfigure[]{\label{subfigure:rrho17}\includegraphics[scale=0.54]{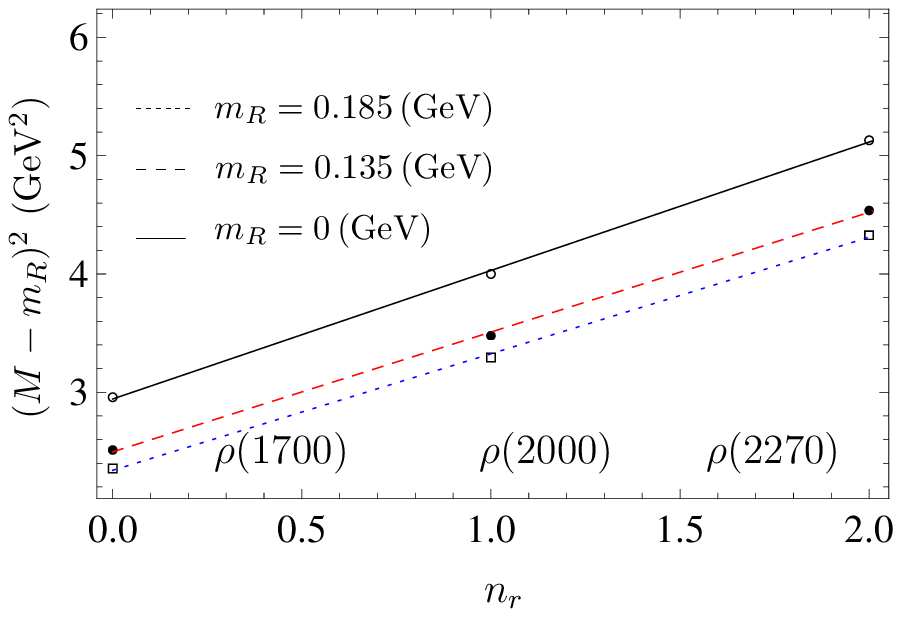}}
\subfigure[]{\label{subfigure:rpi2}\includegraphics[scale=0.54]{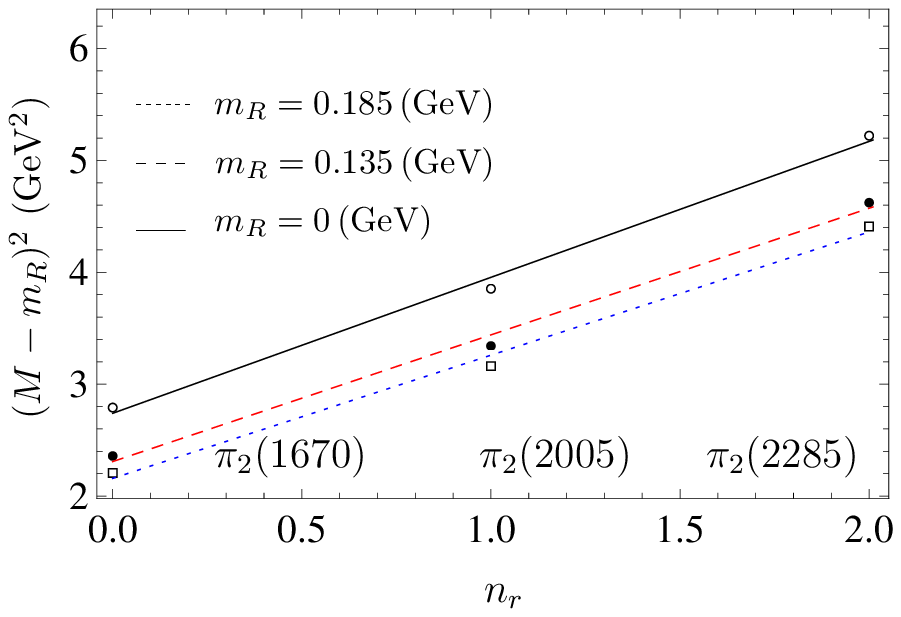}}
\subfigure[]{\label{subfigure:romega3}\includegraphics[scale=0.54]{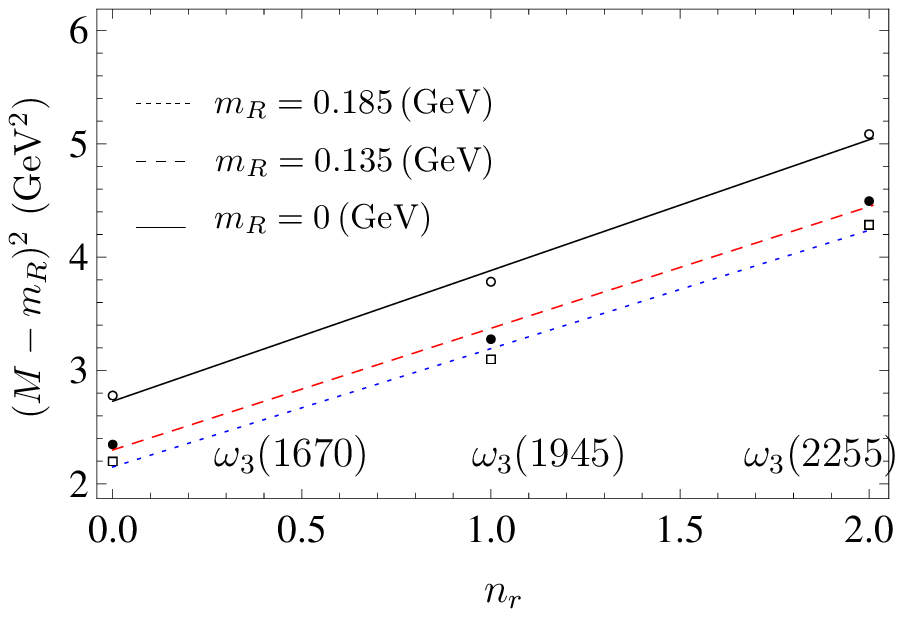}}
\subfigure[]{\label{subfigure:reta2}\includegraphics[scale=0.54]{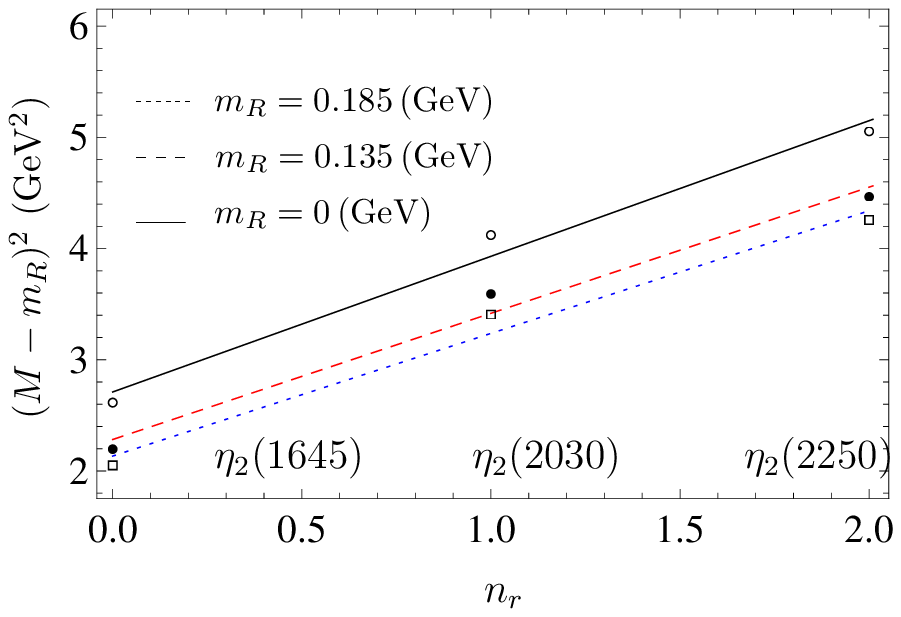}}
\caption{The radial {\rts} for the light nonstrange mesons. They are plotted in the $((M-m_R)^2,\,n_r)$ plane. The used formulas are listed in Table \ref{tab:regmr2}. The experimental masses are from PDG \cite{ParticleDataGroup:2020ssz}.}\label{fig:reggemr2}
\end{figure*}

\begin{figure*}[!phtb]
\centering
\subfigure[]{\label{subfigure:opi0}\includegraphics[scale=0.54]{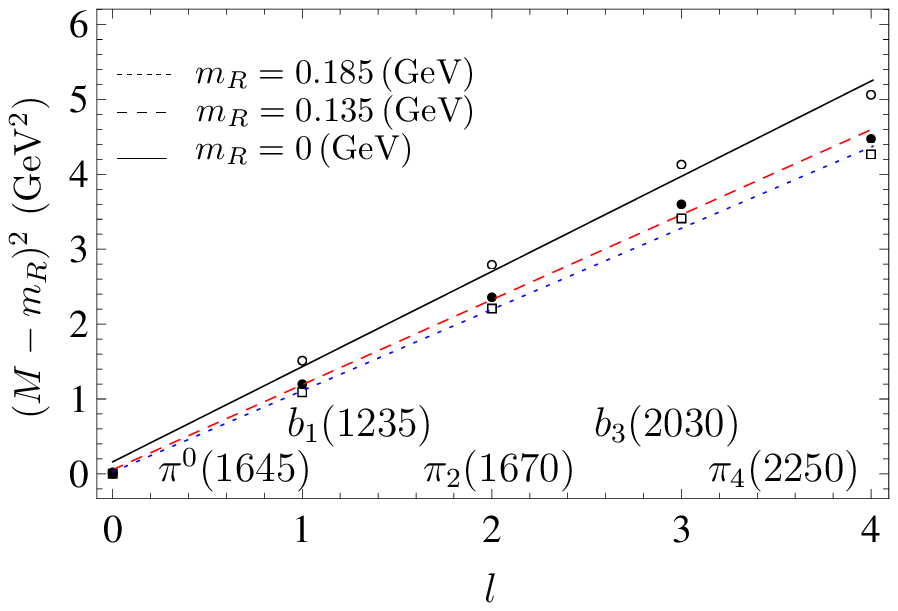}}
\subfigure[]{\label{subfigure:orho7}\includegraphics[scale=0.54]{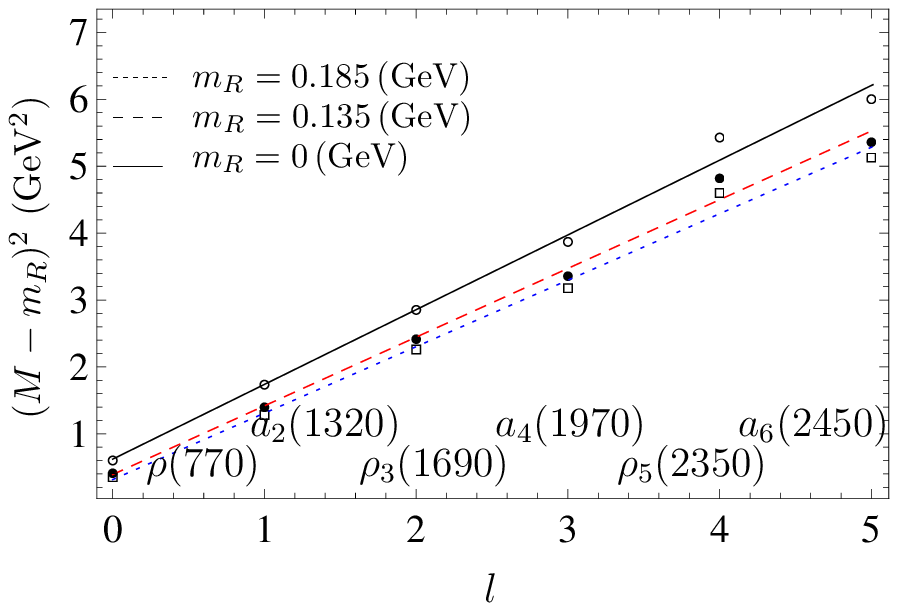}}
\subfigure[]{\label{subfigure:oa112}\includegraphics[scale=0.54]{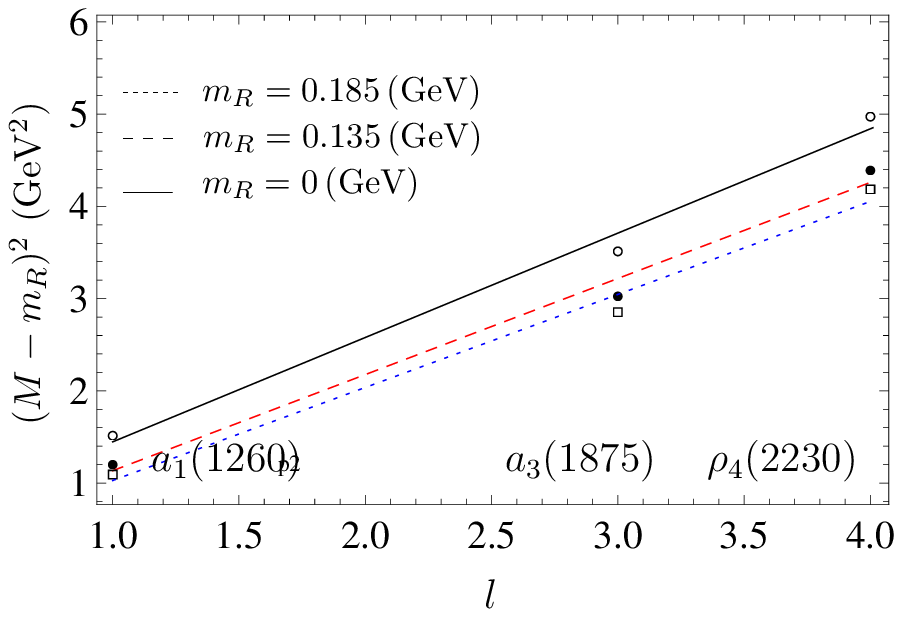}}
\subfigure[]{\label{subfigure:oeta}\includegraphics[scale=0.54]{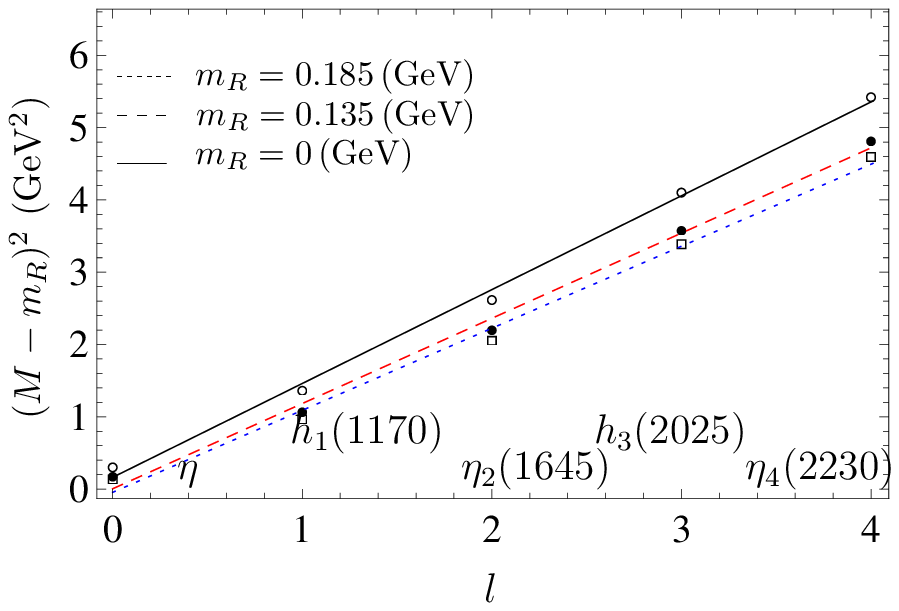}}
\subfigure[]{\label{subfigure:oomega7}\includegraphics[scale=0.54]{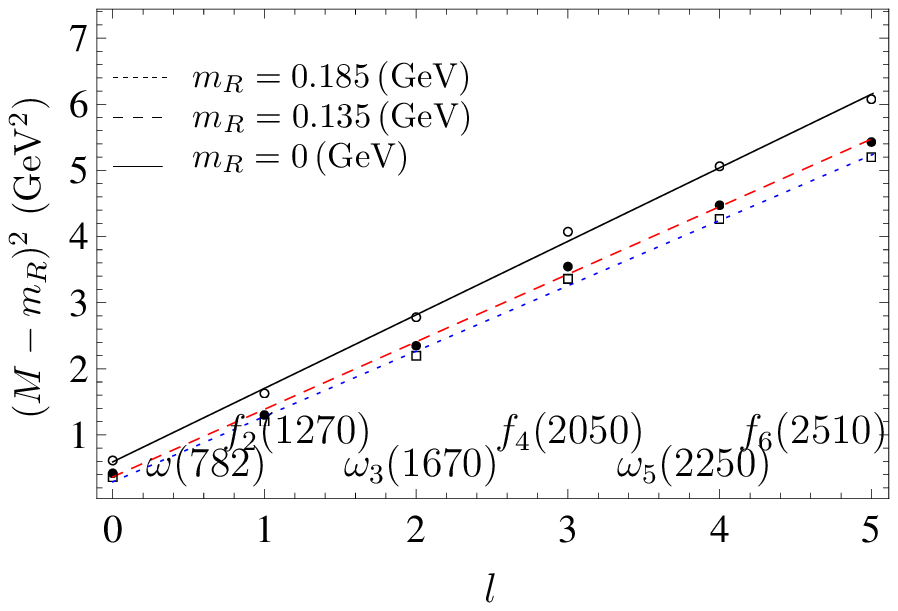}}
\subfigure[]{\label{subfigure:opi13}\includegraphics[scale=0.54]{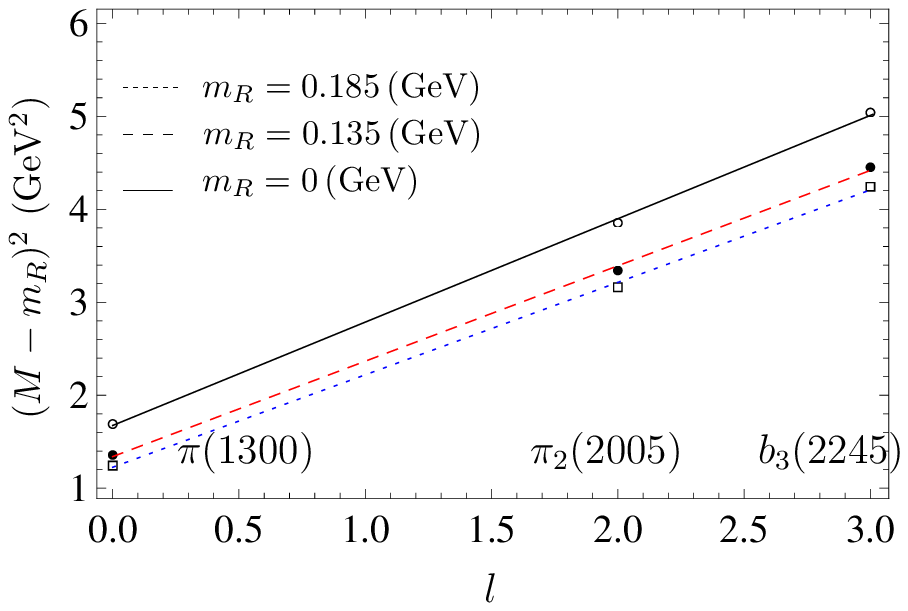}}
\subfigure[]{\label{subfigure:orho14}\includegraphics[scale=0.54]{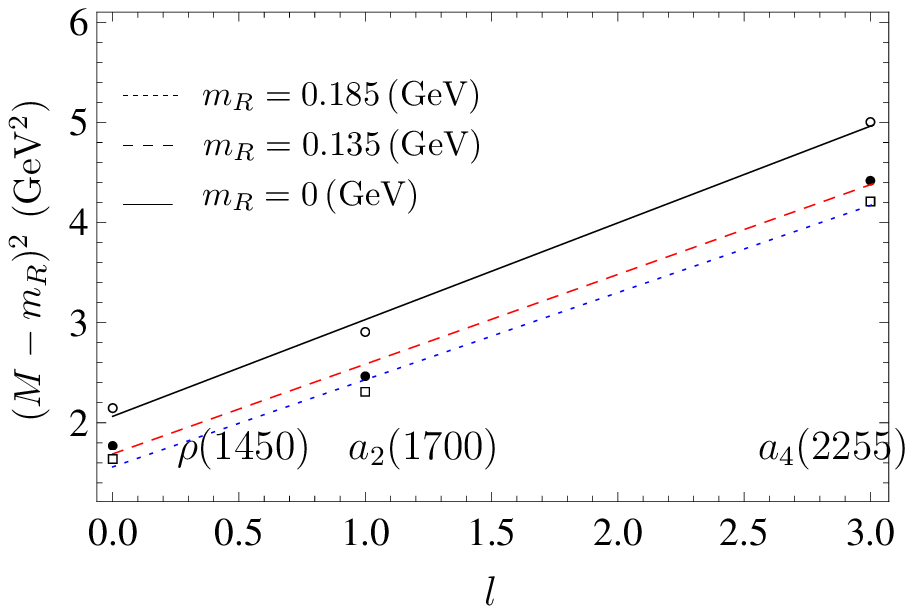}}
\subfigure[]{\label{subfigure:oa116}\includegraphics[scale=0.54]{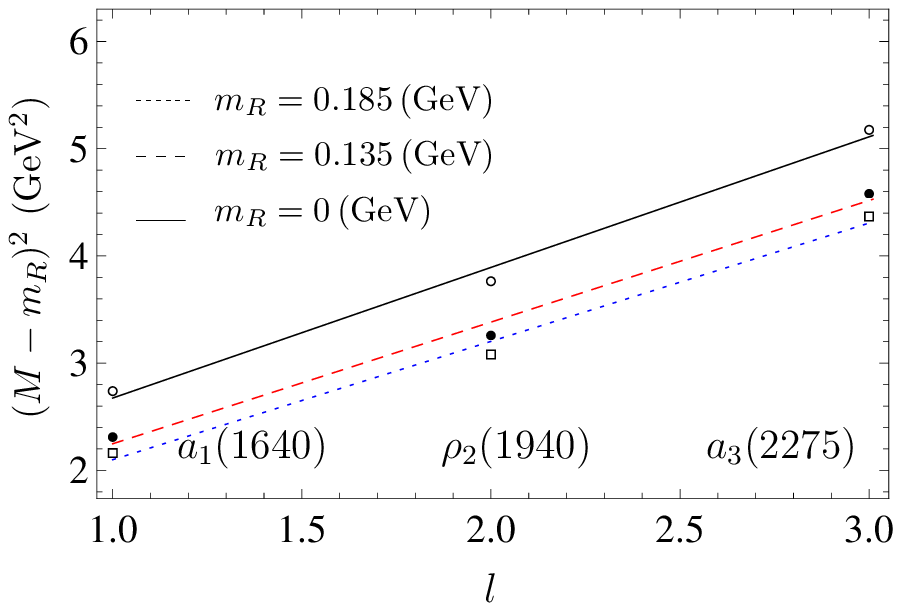}}
\subfigure[]{\label{subfigure:oomega14}\includegraphics[scale=0.54]{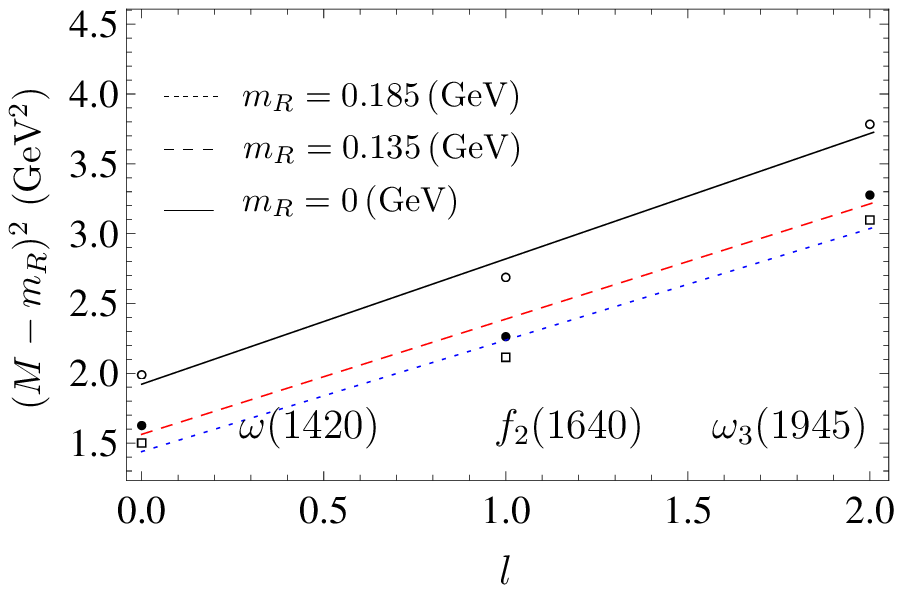}}
\subfigure[]{\label{subfigure:opi18}\includegraphics[scale=0.54]{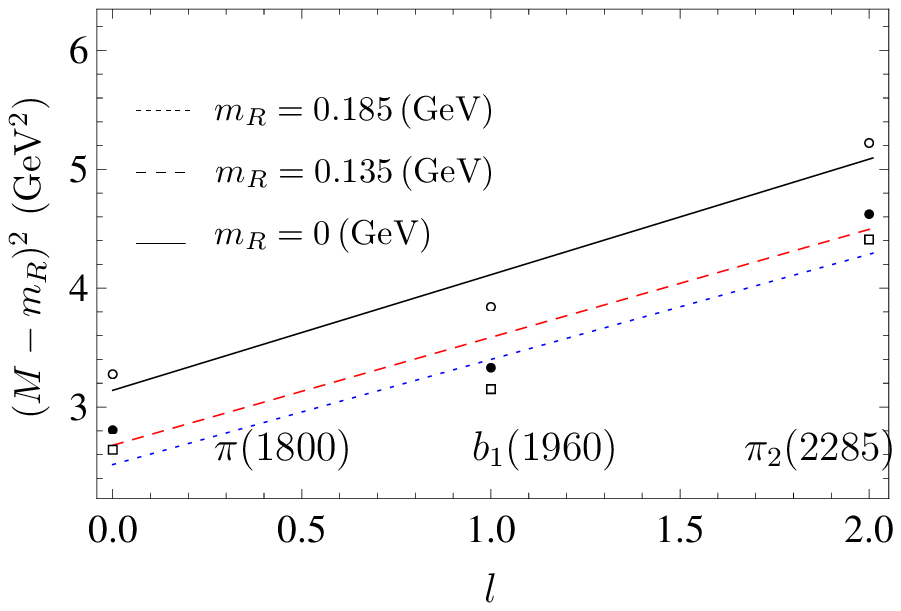}}
\subfigure[]{\label{subfigure:oeta17}\includegraphics[scale=0.54]{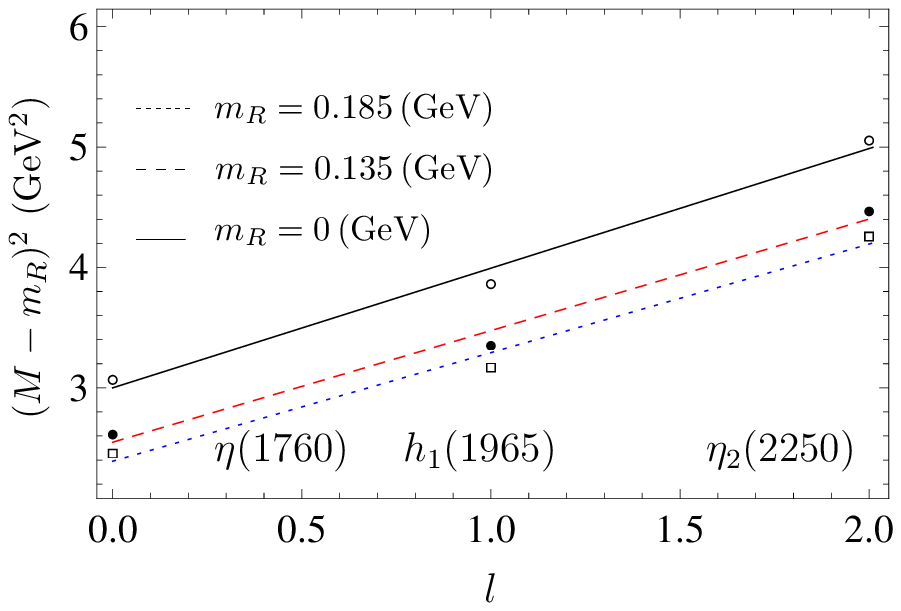}}
\subfigure[]{\label{subfigure:oomega16}\includegraphics[scale=0.54]{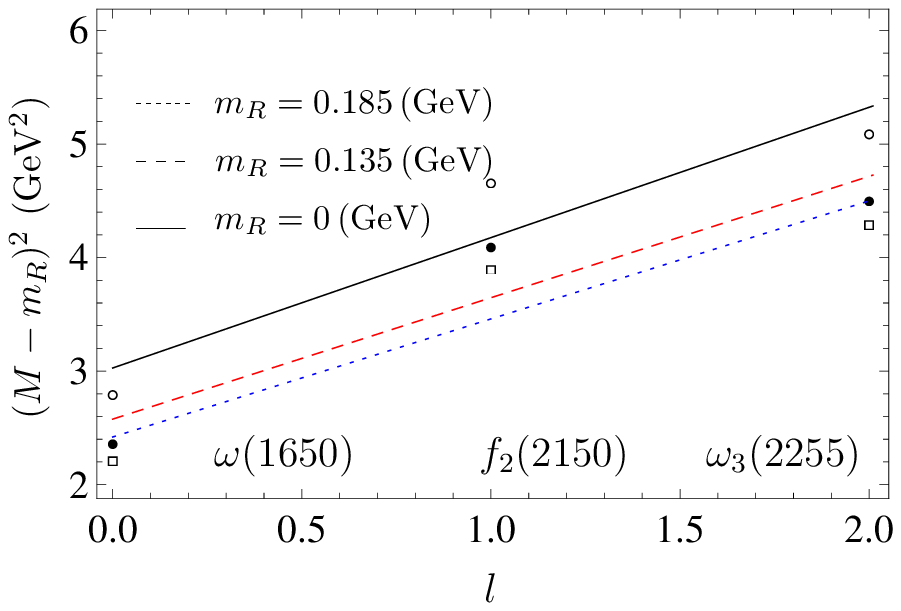}}
\caption{The orbital {\rts} for the light nonstrange mesons. They are plotted in the $((M-m_R)^2,\,l)$ plane. The used formulas are listed in Table \ref{tab:regmr2}. The experimental masses are from PDG \cite{ParticleDataGroup:2020ssz}.}\label{fig:reggemo2}
\end{figure*}

In this section, the {\rts} for the light nonstrange mesons are fitted individually by the formula in Eq. (\ref{reglike}) with $\gamma=1$. The experimental masses are from PDG \cite{ParticleDataGroup:2020ssz}. The fitted {\rts} are listed in Table \ref{tab:regmr2} and shown in Figs. \ref{fig:reggemr2}, \ref{fig:reggemo2}. The {\rt} formed by $a_0(1450)$, $\rho(1700)$ and $\rho_3(1990)$ [$M^2=0.57\,l+2.70$] and the {\rt} formed by $f_0(1370)$, $\omega(1650)$ and $f_2(1810)$ [$M^2=0.74\,l+1.16$] are not listed in Table \ref{tab:regmr2} due to their too small slopes.

As $m_R$ increases, $\alpha_x$ and $\alpha_x{c_0}$ will decrease, see Figs. \ref{fig:reggemr2}, \ref{fig:reggemo2} and Table \ref{tab:regmr2}. The averaged slope $\overline{\alpha}_{n_r}$ for the radial {\rts} varies from $1.24$ $\rm{GeV}^2$ to $1.14$ $\rm{GeV}^2$ and $1.10$ $\rm{GeV}^2$ as $m_R$ is from $0$ $\rm{GeV}$ to $0.135$ $\rm{GeV}$ and $0.185$ $\rm{GeV}$, see Table \ref{tab:regmr2alpha}. The averaged slope $\overline{\alpha}_{l}$ for the orbital {\rts} is $1.10$ $\rm{GeV}^2$, $1.02$ $\rm{GeV}^2$ and $0.98$ $\rm{GeV}^2$ for $m_R=0$ $\rm{GeV}$, $m_R=0.135$ $\rm{GeV}$ and $m_R=0.185$ $\rm{GeV}$, respectively.

For the conventional form of the Regge trajectories $M^2=\alpha_l l+\alpha_{n_r} n_r +c$, $\alpha_l$ is not always equal to $\alpha_{n_r}$, see Table \ref{tab:regmr2}. If $x \equiv l+ {\alpha_{n_r}/\alpha_l} n_r$, the averaged slopes $\overline{\alpha}_l$ is not equal to the averaged slopes $\overline{\alpha}_{n_r}$ for the light nonstrange mesons, see Table \ref{tab:regmr2alpha}. The ratio ${\alpha_{n_r}/\alpha_l}$ lies in the interval $0.85\le\alpha_{n_r}/\alpha_l\le1.28$ as $l=n_r=0$. As $l=n_r=1$, $0.93\le\alpha_{n_r}/\alpha_l\le1.68$. As $l=n_r=2$, $0.95\le\alpha_{n_r}/\alpha_l\le1.26$. The obtained results are consistent with Refs. \cite{Brau:2000st,Kang:1975cq,Afonin:2007jd}. The effect of $m_R$ on the ratio $\alpha_{n_r}/\alpha_l$ is small as $m_R$ ranges from $0$ $\rm{GeV}$ to $0.185$ $\rm{GeV}$.

\begin{table*}[!phtb]
\caption{The linearly fitted {\rts} for the light nonstrange mesons in the $((M-m_R)^2,\,x)$ plane, where $x=n_r,\,l$. }
\centering
\begin{tabular*}{\textwidth}{@{\extracolsep{\fill}}cclll@{}}
\hline\hline
  Traj.  &   &  $m_R=0$ {\rm GeV}  &   $m_R=0.135$ {\rm GeV}  & $m_R=0.185$ {\rm GeV}   \\
\hline
  $\pi^0$  & $l=0$  & 1.37 $n_r$ + 0.23   &  1.23 $n_r$ + 0.11     & 1.18 $n_r$ + 0.08   \\
$\rho(770)$  & $l=0$ & 1.42 $n_r$ + 0.67   & 1.30 $n_r$ + 0.46      & 1.25 $n_r$ + 0.38  \\
  $\eta$     & $l=0$  & 1.25 $n_r$ + 0.39   & 1.14 $n_r$ + 0.21     & 1.10 $n_r$ + 0.15 \\
  $\omega(782)$  & $l=0$    & 1.11 $n_r$ + 0.67   & 1.02 $n_r$ + 0.44     & 0.98 $n_r$ + 0.37  \\
%
  $a_1(1260)$  & $l=1$     & 1.26 $n_r$ + 1.56   & 1.16 $n_r$ + 1.24      & 1.13 $n_r$ + 1.12  \\
  $a_2(1320)$  & $l=1$      & 1.14 $n_r$ + 1.75   & 1.05 $n_r$ + 1.40      & 1.01 $n_r$ + 1.29   \\
  $b_1(1235)$  & $l=1$      & 1.17 $n_r$ + 1.51   & 1.08 $n_r$ + 1.19      & 1.04 $n_r$ + 1.08   \\
$f_2(1270)$   & $l=1$  & 1.51 $n_r$ + 1.48   & 1.39 $n_r$ + 1.16     & 1.35 $n_r$ + 1.05  \\
  $h_1(1170)$ & $l=1$      & 1.19 $n_r$ + 1.38   & 1.10 $n_r$ + 1.09     & 1.06 $n_r$ + 0.98  \\
%
$\rho(1700)$  & $l=2$  & 1.09 $n_r$ + 2.94   & 1.01 $n_r$ + 2.50     & 0.99 $n_r$ + 2.34   \\
  $\pi_2(1670)$  & $l=2$     & 1.22 $n_r$ + 2.74   & 1.13 $n_r$ + 2.31     & 1.10 $n_r$ + 2.16   \\
  $\omega_3(1670)$  & $l=2$  & 1.15 $n_r$ + 2.73  & 1.07 $n_r$ + 2.30      & 1.04 $n_r$ + 2.15  \\
  $\eta_2(1645)$  & $l=2$    & 1.22 $n_r$ + 2.71  & 1.13 $n_r$ + 2.28      & 1.10 $n_r$ + 2.13  \\
\hline
$\pi^0/b_1$    & $n_r=0$      & 1.27 $l$ + 0.16   & 1.13 $l$ + 0.06         & 1.08 $l$ + 0.03 \\
$\rho(770)/a_2$  & $n_r=0$   & 1.12 $l$ + 0.62    &   1.03 $l$ + 0.39       & 0.99 $l$ + 0.32       \\
$a_1(1260)/\rho_2$  & $n_r=0$  &1.13 $l$ + 0.32   &1.04 $l$ + 0.09      &1.01 $l$ + 0.02  \\
  $\eta/h_1$   & $n_r=0$  & 1.30 $l$ + 0.16    &   1.18 $l$ + 0.       & 1.13 $l$ - 0.04       \\
  $\omega(782)/f_2$  & $n_r=0$  & 1.11 $l$ + 0.59   &   1.02 $l$ + 0.36       & 0.99 $l$ + 0.29       \\
%
%
%
$\pi(1300)/b_1$  & $n_r=1$   & 1.11 $l$ + 1.68   &1.03 $l$ + 1.34        & 0.99 $l$ + 1.23    \\
$\rho(1450)/a_2$  & $n_r=1$  & 0.97 $l$ + 2.06   &0.90 $l$ + 1.69        &0.87 $l$ + 1.56  \\
$a_1(1640)/\rho_2$  & $n_r=1$  &1.22 $l$ + 1.46  &1.13 $l$ + 1.11        &1.10 $l$ + 1.00  \\
$\omega(1420)/f_2$  & $n_r=1$  &0.90 $l$ + 1.92  &0.83 $l$ + 1.56        &0.80 $l$ + 1.44 \\
$\pi(1800)/b_1$  & $n_r=2$   &0.97 $l$ + 3.14   &0.91 $l$ + 2.68        & 0.88 $l$ + 2.52 \\
$\eta(1760)/h_1$  & $n_r=2$  &0.99 $l$ + 3.00   &0.93 $l$ + 2.55        & 0.90 $l$ + 2.39 \\
$\omega(1650)/f_2$  & $n_r=2$  &1.15 $l$ + 3.03 &1.07 $l$ + 2.58        &1.04 $l$ + 2.42\\
\hline\hline
\end{tabular*}\label{tab:regmr2}
\end{table*}

\begin{table*}[!phtb]
\caption{The averaged slopes of the fitted {\rts} for the light nonstrange mesons in the $((M-m_R)^2,\,x)$ plane, where $x=n_r,\,l$. The used data are from Table \ref{tab:regmr2}. $\overline{\alpha}_{n_r}$ and  $\overline{\alpha}_l$ are in units of $\rm GeV^2$.}
\centering
\begin{tabular*}{\textwidth}{@{\extracolsep{\fill}}llccc@{}}
\hline\hline
    &    &    $m_R=0$ $\rm GeV$   &   $m_R=0.135$ $\rm GeV$  & $m_R=0.185$ $\rm GeV$   \\
\hline
                      &     $l=0$    & 1.29   &  1.17      & 1.13   \\
$\overline{\alpha}_{n_r}$&  $l=1$    & 1.25  &  1.16      & 1.12   \\
                         &  $l=2$    & 1.17  &  1.09      & 1.07    \\
                         &  $l=0,1,2$  & 1.24  &  1.14     &  1.10    \\
                        &  $n_r=0$   & 1.19   &  1.08      & 1.04 \\
  $\overline{\alpha}_l$ &  $n_r=1$    & 1.05   &  0.97      & 0.94    \\
                        &  $n_r=2$    & 1.04   &  0.97     &  0.94   \\
                        &  $n_r=0,1,2$  & 1.10   & 1.02     &   0.98  \\
\hline\hline
\end{tabular*}\label{tab:regmr2alpha}
\end{table*}

\end{document}